\documentclass[a4paper,fleqn,usenatbib]{mnras}

\usepackage[T1]{fontenc}
\usepackage[utf8]{inputenc}
\usepackage{ae,aecompl}

\usepackage{graphicx}	
\usepackage{amsmath}	
\usepackage{amssymb}	
\usepackage{booktabs}
\usepackage{longtable}

\usepackage{txfonts}

\usepackage{eso-pic}

\AddToShipoutPictureBG*{%
  \AtPageUpperLeft{%
    \hspace{0.75\paperwidth}%
    \raisebox{-3.5\baselineskip}{%
        \makebox[0pt][l]{\textnormal{DES-2018-0391}}
}}}%

\AddToShipoutPictureBG*{%
  \AtPageUpperLeft{%
    \hspace{0.75\paperwidth}%
    \raisebox{-4.5\baselineskip}{%
        \makebox[0pt][l]{\textnormal{FERMILAB-PUB-18-552-AE}}
}}}%

\title[ConvNets and high-z DES lenses]{Finding high-redshift strong lenses in DES using convolutional neural
networks}
\author[C. Jacobs, T. Collett, K. Glazebrook, C. McCarthy, K. Qin et al]{
\parbox{\textwidth}{
\Large
C.~Jacobs$^{1,2}$\thanks{E-mail:colinjacobs@swin.edu.au},
T.~Collett$^{3}$,
K.~Glazebrook$^{1,2}$,
C.~McCarthy$^{4}$,
A.K.~Qin$^{4}$,
T.~M.~C.~Abbott$^{5}$,
F.~B.~Abdalla$^{6,7}$,
J.~Annis$^{8}$,
S.~Avila$^{3}$,
K.~Bechtol$^{9}$,
E.~Bertin$^{10,11}$,
D.~Brooks$^{6}$,
E.~Buckley-Geer$^{8}$,
D.~L.~Burke$^{12,13}$,
A.~Carnero~Rosell$^{14,15}$,
M.~Carrasco~Kind$^{16,17}$,
J.~Carretero$^{18}$,
L.~N.~da Costa$^{14,15}$,
C.~Davis$^{12}$,
J.~De~Vicente$^{19}$,
S.~Desai$^{20}$,
H.~T.~Diehl$^{8}$,
P.~Doel$^{6}$,
T.~F.~Eifler$^{21,22}$,
B.~Flaugher$^{8}$,
J.~Frieman$^{8,23}$,
J.~Garc\'ia-Bellido$^{24}$,
E.~Gaztanaga$^{25,26}$,
D.~W.~Gerdes$^{27,28}$,
D.~A.~Goldstein$^{29,30}$,
D.~Gruen$^{12,13}$,
R.~A.~Gruendl$^{16,17}$,
J.~Gschwend$^{14,15}$,
G.~Gutierrez$^{8}$,
W.~G.~Hartley$^{6,31}$,
D.~L.~Hollowood$^{32}$,
K.~Honscheid$^{33,34}$,
B.~Hoyle$^{35,36}$,
D.~J.~James$^{37}$,
K.~Kuehn$^{38}$,
N.~Kuropatkin$^{8}$,
O.~Lahav$^{6}$,
T.~S.~Li$^{8,23}$,
M.~Lima$^{39,14}$,
H.~Lin$^{8}$,
M.~A.~G.~Maia$^{14,15}$,
P.~Martini$^{33,40}$,
C.~J.~Miller$^{27,28}$,
R.~Miquel$^{41,18}$,
B.~Nord$^{8}$,
A.~A.~Plazas$^{22}$,
E.~Sanchez$^{19}$,
V.~Scarpine$^{8}$,
M.~Schubnell$^{28}$,
S.~Serrano$^{25,26}$,
I.~Sevilla-Noarbe$^{19}$,
M.~Smith$^{42}$,
M.~Soares-Santos$^{43}$,
F.~Sobreira$^{44,14}$,
E.~Suchyta$^{45}$,
M.~E.~C.~Swanson$^{17}$,
G.~Tarle$^{28}$,
V.~Vikram$^{46}$,
A.~R.~Walker$^{5}$,
Y.~Zhang$^{8}$,
J.~Zuntz$^{47}$,
\begin{center} (DES Collaboration) \end{center}
}
\vspace{0.4cm}
\\
\parbox{\textwidth}{
Affiliations listed at the end of this document
}
}

\begin{document}

\date{}
\pagerange{\pageref{firstpage}--\pageref{lastpage}} \pubyear{2016}

\maketitle

\label{firstpage}

 \begin{abstract}
We search Dark Energy Survey (DES) Year 3 imaging data for galaxy-galaxy
strong gravitational lenses using convolutional neural networks. We
generate 250,000 simulated lenses at redshifts > 0.8 from which
we create a data set for training the neural networks with
realistic seeing, sky and shot noise. Using the simulations as a guide,
we build a catalogue of 1.1 million DES sources with
\(1.8 < g - i < 5\), \(0.6 < g -r < 3\), r\_mag \textgreater{} 19,
g\_mag \textgreater{} 20 and i\_mag \textgreater{} 18.2. We train two
ensembles of neural networks on training sets consisting of simulated
lenses, simulated non-lenses, and real sources. We use the neural
networks to score images of each of the sources in our catalogue with a
value from 0 to 1, and select those with scores greater than a chosen
threshold for visual inspection, resulting in a candidate set of 7,301
galaxies. During visual inspection we rate 84 as "probably" or "definitely"
lenses. Four of these are previously known lenses or lens candidates. We
inspect a further 9,428 candidates with a different score threshold, and
identify four new candidates. We present 84 new strong lens candidates,
selected after a few hours of visual inspection by astronomers. 
This catalogue contains a comparable number of high-redshift lenses
to that predicted by simulations. Based on
simulations we estimate our sample to contain most discoverable 
lenses in this imaging and at this redshift range.
\end{abstract}

\begin{keywords}
gravitational lensing: strong -- methods: statistical
\end{keywords}

\section{Introduction}\label{introduction}

Gravitational lensing, a phenomenon arising from the relativistic
curvature of spacetime around massive objects
\citep{EinsteinLenslikeactionstar1936, ZwickyNebulaeGravitationalLenses1937},
is a subject of increasing importance in astrophysics and cosmology.
Where a large lensing potential and a close alignment of
the lens mass and source coincide, strong lensing can produce highly magnified images
of distant sources. When studied, they can serve as a unique probe of both
lens and source properties \citep[see][ for an
overview]{treu_strong_2010}. Since the detection of the first strongly
lensed quasar in 1979 \citep{Walsh0957561twin1979} a growing catalogue
of strong lenses has been discovered, now numbering in the
hundreds\footnote{L.A. Moustakas \& J. Brownstein, priv. comm. Database
  of confirmed and probable lenses from all sources, curated by the
  University of Utah. http://admin.masterlens.org}.

Individual strong lenses can be highly valuable scientifically. By
magnifying distant sources by a factor of tens to \textasciitilde{}100,
lensing can allow us to examine sources otherwise too distant to detect,
for instance
\citep{Starkformationassemblytypical2008, Quiderultravioletspectrumgravitationally2009, NewtonSloanLensACS2011, Zhengmagnifiedyounggalaxy2012, EbelingThirtyfoldExtremeGravitational2018},
even a single star at redshift 1.5
\citep{Kellyindividualstarredshift2017}. In quantity, strong lenses can
be valuable cosmological probes; the many applications include an
independent measure of \(\mbox{H}_0\) via time delays between
multiply-imaged quasars \citep{BonvinH0LiCOWNewCOSMOGRAIL2016}, or
testing Warm Dark Matter models through the statistics of perturbations
in a large sample of Einstein rings and arcs
\citep{VegettiGravitationaldetectionlowmass2012, LiConstraintsidentitydark2016},
including by line-of-sight substructure 
\citep{DespaliModellinglineofsightcontribution2018}.
For the latter, lenses at high redshift are particularly valuable.

Because of their high surface mass density, Early Type Galaxies
(ETGs) represent the vast majority of galaxy-galaxy lenses. ETGs contain
most of the stellar mass in the local universe, and so an understanding
of their star formation and assembly histories is key for building an
accurate picture of the evolution of structure in the universe.
Strong lensing can act as a probe of lens mass with precision at great
distances, and is thus a crucial tool in understanding the history of
these galaxies at early times.

Observations have shown that the total density profiles of elliptical
galaxies can be well-described by a power law, with
\(\rho(r) \propto r^{-\gamma'}\). Observationally, most galaxies
demonstrate roughly isothermal profiles, i.e. \(\gamma' \sim 2\),
however reproducing the observed isothermality has proven challenging
for simulations. Magneticum and EAGLE simulations both predict slopes
significantly shallower than observed in local galaxies
\citep{BellstedtSLUGGSSurveycomparison2018a}. Simulations also predict
that \(\gamma'\) becomes shallower over time
\citep{RemusCoEvolutionTotalDensity2017}, whilst observations suggest
the opposite
\citep{SonnenfeldSL2SGalaxyscaleLens2013, ShankarRevisitingbulgehalo2018}.
This tension implies that our understanding of the mechanisms by which
galaxies evolve, such as the role of dissipationless dry mergers at
later times, is incomplete. At the present time, the redshift leverage
of existing observations is insufficient to settle this question; only
five lenses at redshift \(>0.8\) have been available for this analysis.

Locally, the mass density profiles of ETGs have been probed using tools
such as stellar dynamics
\citep[notably][]{TimdeZeeuwSAURONprojectII2002, CappellariATLAS3Dprojectvolumelimited2011}
and the dynamics of HI gas regions
\citep[e.g.][]{Weijmansshapedarkmatter2008} and globular clusters
\citep[e.g.][]{OldhamGalaxystructuremultiple2018}, however beyond the
local universe lensing is the most practical tool. The Einstein radius
of a lens system is an observable quantity and is proportional to the
mass within that radius; combined with a measurement of velocity
dispersion and source and lens redshifts, a robust measurement of the
Einstein radius can constrain \(\gamma'\), the mean total density slope,
to under five percent
\citep{TreuMassiveDarkMatter2004, treu_strong_2010, RuffSL2SGalaxyscaleLens2011}.
This analysis has been carried out at local redshifts, for instance by
\citet{CollierImprovedmassconstraints2018} (two galaxies at z=0.03 and
z=0.05); on 16 Sloan Lens ACS Survey (SLACS) galaxies in the redshift 
range 0.08 - 0.33 by \citet{BarnabeTwodimensionalkinematicsSLACS2011};
and on 25 Strong Lensing Legacy Survey (SL2S) galaxies at
redshifts 0.2 - 0.8 by \citet{SonnenfeldSL2SGalaxyscaleLens2013},
constraining \(\gamma'\) to \textasciitilde{}5\% in that range. A bigger
sample of lenses at redshift \(> 0.8\) is needed to confirm the
evolution of gamma with redshift and thereby constrain simulations and
our corresponding understanding of the physics of galaxy evolution.

Finding strong lenses, especially at higher redshifts, remains a
significant challenge. Currently several hundred examples of confirmed
or likely galaxy-galaxy strong lenses have been discovered \citep[the
Masterlens database\footnote{L.A. Moustakas \& J. Brownstein, priv.
  comm. Database of confirmed and probable lenses from all sources,
  curated by the University of Utah. http://admin.masterlens.org}]{collett_population_2015},
with several hundred more awaiting spectroscopic or high-resolution
follow up. Modelling such as \citep{collett_population_2015} and
\citep{treu_strong_2010} predicts that several thousand lenses should be
detectable in current surveys such as the Dark Energy Survey
\citep[DES;][]{TheDESCollaborationDarkEnergySurvey2005} and tens of
thousands in next-generation surveys such as the Large Synoptic Survey
Telescope \citep[LSST;][]{IvezicLSSTScienceDrivers2008} and Euclid
\citep{AmiauxEuclidMissionbuilding2012}. 

In the past, entire surveys
could be searched by eye, but the data sets are now of a scale that
makes this impractical. Previous strategies for automating the lens
search have included searching images for characteristic features such
as arcs and rings
\citep{LenzenAutomaticdetectionarcs2004, AlardAutomateddetectiongravitational2006, EstradaSystematicSearchHigh2007, seidel_arcfinder:_2007, MoreCFHTLSStrongLensing2012, gavazzi_ringfinder:_2014},
searching for red-near-blue sources
\citep{BoltonSloanLensACS2006, DiehlBrightArcsSurvey2017}, applying machine learning
to survey catalogs \citep{AgnelloDatamininggravitationally2015}, and modelling
sources as lenses and testing the quality of the residual for a match
\citep{marshall_automated_2009, ChanChitahStronggravitationallensHunter2015}.
Citizen scientists have also been recruited, with 30,000 volunteers
helping to search the Canada-France-Hawaii Telescope Legacy Survey
(CFHTLS) for strong lenses
\citep{MarshallSPACEWARPSCrowdsourcing2016, MoreSPACEWARPSII2016}. Some
recent efforts have focused on machine learning techniques, in
particular ``Deep Learning'', involving the use of large
Artificial Neural Networks. These techniques have already proved
effective at finding lenses. Neural nets can effectively distinguish
between simulated lenses and non-lenses
\citep{JacobsFindingstronglenses2017, LanusseCMUDeepLensDeep2017, AvestruzAutomatedLensingLearner2017, HezavehFastAutomatedAnalysis2017}.
Applying the technique to surveys, 
Jacobs et al \citeyearpar{JacobsFindingstronglenses2017} used an
ensemble of CNNs to find several hundred previously known lenses and 17
new candidates in CFHTLS in under an hour of astronomer review time, and
Petrillo et al
\citeyearpar{PetrilloFindingStrongGravitational2017} used CNNs to
identify 56 new lens candidates in the Kilo Degree Survey (KiDS).

In DES, previous searches have relied heavily on the inspection of many thousands
of candidates chosen from catalogue photometry; see 
section~\ref{comparison-to-other-des-strong-lens-searches}. Collett's 
\citeyearpar{collett_population_2015} simulation
suggests that approximately 8\% of detectable lenses (\({\sim}110\)
lenses) should lie at redshifts \(> 0.8\). It is these lenses that are
the target of the search detailed in this work.

In this paper we describe a first search for high-redshift lenses in the
Dark Energy Survey using machine learning techniques. The paper is
structured as follows: In section~\ref{sec:background}, we provide some
brief background on the machine learning technique employed in the
search, namely artificial neural networks. In section~\ref{sec:method}
we outline the methodology for constructing simulations to train the
neural networks, building a catalogue of sources to search, and
employing the trained networks on survey data. In
section~\ref{sec:results} we present the results of the search. In
section~\ref{sec:discussion} we consider ways to evaluate the
performance of the lens-finding method and improve future searches, and
some prospects for follow-up science and further development of the
technique, the summarise our conclusions in
section~\ref{sec:conclusions}.

\begin{center}\rule{0.5\linewidth}{\linethickness}\end{center}

\section{Artificial Neural Networks}\label{sec:background}

\begin{figure*}
\centering
\includegraphics[width=0.95000\textwidth]{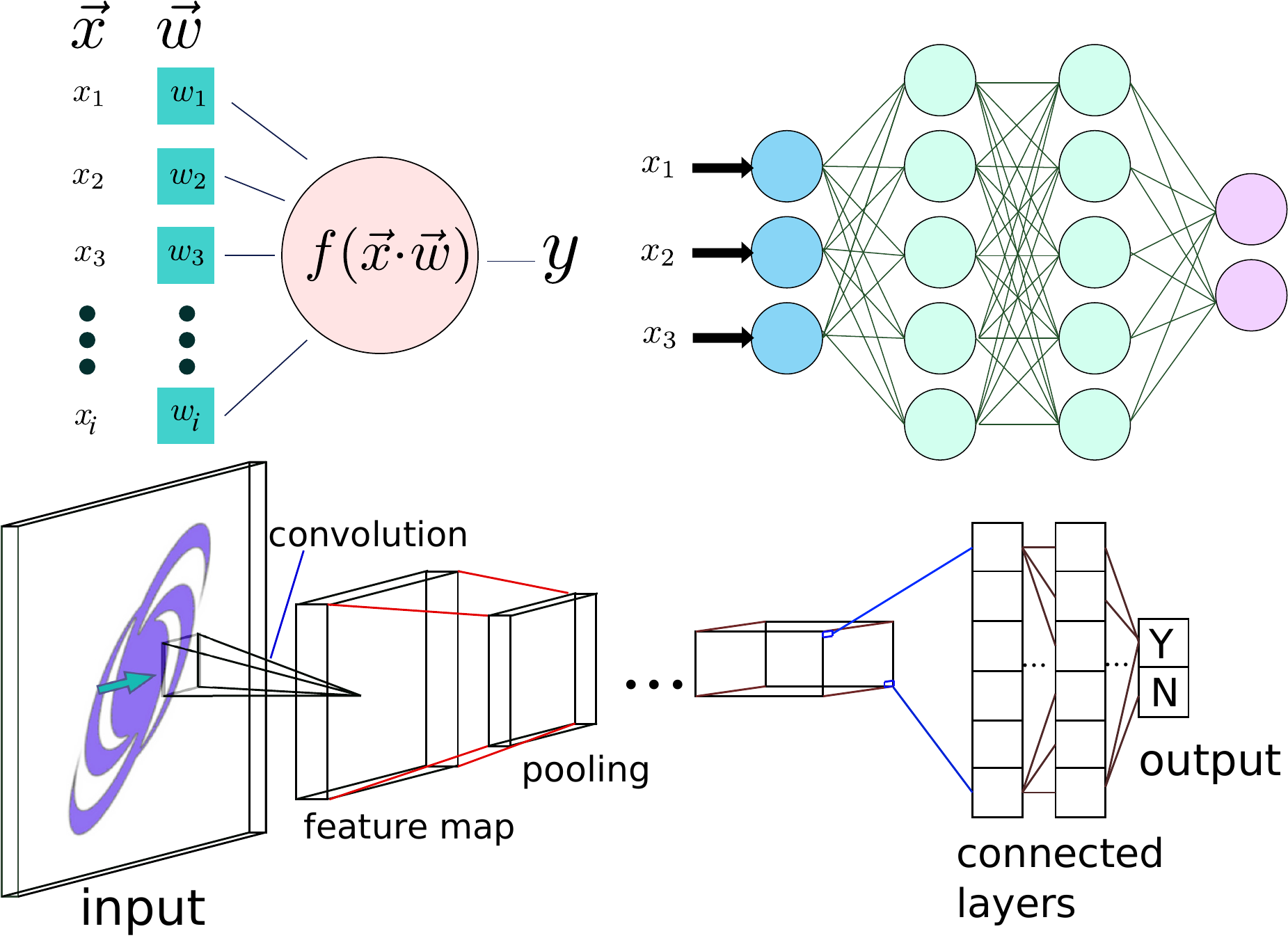}
\caption{Overview of artificial networks. \emph{Top left}: An artificial
neuron. The elementwise weighted sum of a vector input is passed through
a nonlinear activation function (such as the logistic function,
arctangent, or rectified linear unit) to produce a scalar output.
\emph{Top right}: An example of a small fully-connected ANN, consisting of layers of
artificial neurons. Blue: Input layer, equal in size to dimensionality
of input data. Green: Hidden layers. Purple: Output layer; the outputs
are interpreted according to the problem domain. \emph{Bottom}:
Prototypical convolutional neural network. Convolutional kernels are
scanned across the input to build up so-called feature maps; pooling
layers subsample the preceding layer to reduce the spatial extent. This
process is repeated some number of times, then the resulting feature
maps are passed to one or more fully-connected layers, followed by the
output (yes/no) as the last layer.}\label{fig:ann}

\end{figure*}

Here we employ a machine learning technique to automatically find
galaxy-galaxy strong lenses in DES image data. While traditional
approaches to data problems rely on algorithms developed by subject-matter 
experts who define key features in the data and their relative
contributions to the problem space, machine learning techniques extract
features and their importance from data alone. See
\citep{JordanMachinelearningtrends2015} for an overview of the theory
and applications of machine learning. Artificial Neural Networks (ANNs)
are a machine learning technique first developed in the 1950s
\citep{RosenblattPerceptronaperceivingrecognizing1957} and more heavily
researched in the 1980s and 1990s
\citep{FukushimaNeocognitronselforganizingneural1980} as non-trivial
networks became computationally more tractable. ANNs are constructed to
loosely mimic the structure of the brain, with a network of
interconnected `neurons', the strengths of the connection influencing
how each neuron responds to a signal from its peers. Each artificial
neuron takes an input vector; calculates the dot product with a vector
of weights (i.e.~real numbers that weight the contribution of each input
value); and passes the resulting scalar through a non-linear function
such as a logistic function or hyperbolic tangent. Neurons are arranged
in layers, with an input layer at one end, an arbitrary number of
``hidden layers'' and an output layer interpreted appropriately to the
problem domain, such as the probabilities a given input lies in one of N
classes (see Figure~\ref{fig:ann}). In theory, the connections between
the neurons/layers can represent a highly non-linear decision boundary
in many dimensions. The process of finding optimal values for the
weights - the training - is data-driven (see below).

The combination of improved technique, widely-available GPU computing,
and the availability of large, labelled datasets means that large ANNs
with many layers (``deep'' ANNs) are now practical. This ``Deep
Learning'' resurgence has revolutionised several fields such as computer
vision and speech recognition which were able to make breakthroughs in
accuracy exceeding the performance of the best hand-engineered
algorithms by large margins
\citep{SchmidhuberDeeplearningneural2015, GuoDeeplearningvisual2016}.

Convolutional Neural Networks
\citep[CNNs;][]{LeCunBackpropagationAppliedHandwritten1989} in
particular have proven highly effective at discovering patterns in image
data. Unlike a standard ANN, where each layer is fully connected to the
previous layer, a convolutional layer connects only small groups of
neighbouring neurons, and shares the weights between groups. This has
the effect of vastly reducing the number of trainable weights while at
the same time taking advantage of the fact that in visual data
neighbouring inputs - i.e.~pixels - are highly correlated in meaningful
ways. In effect, the network uses (usually square - e.g.~5x5 pixels)
`convolutional kernels' which are convolved with the input image or
outputs of a previous layer, and act as feature detectors. Outputs are then
pooled, taking the mean or maximum value of groups of pixels, reducing the spatial
extent of the data as the number of feature maps increases. At earlier
layers, raw features such as edges and patches of colour are detected;
at later layers the network detects patterns in an increasingly abstract
and high-level feature space. Thus at early layers the network activates
on lines and curves; at intermediate layers on combinations of these
into semantically meaningful features; then at later layers, combining
these semantic features into a representation of the input in a
classification space.

A CNN large enough to, for instance, distinguish between objects in
hundreds of categories or decipher audio data into speech contains
millions to hundreds of millions of parameters to be trained. This
requires a large (i.e.~up to millions of examples) training set of
labelled data with which to optimise the weights to achieve the desired
output semantics. The full process for training a neural network,
including the backpropagation algorithm, is detailed in
\citet{LeCunEfficientBackProp1998}. In brief, we construct a loss
function \(L\) such that \(L = 0\) if the network classifies the
training set perfectly, and increases as performance accuracy decreases.
A typical loss function, and the one employed here, is a cross entropy
loss function \citep{CaoLearningRankPairwise2007}\footnote{\(H(y, \hat{y}) = - \sum_{i} -y log\hat{y} -(1-y)log(1-\hat{y})\),
where \(y \in {0, 1}\) are the ground-truth categories and \(\hat(y) \) are the predicted 
    probabilities. \(H(y, \hat{y}) = 0\) if \(\hat{y} = y\).}.

For each training example or batch of examples, and for each of the
trainable weights \(w_i\) in the network, we calculate the gradient
\(\delta_i = \partial{L}/\partial{w_i}\). Then, following the standard
gradient descent paradigm, we update the weights by \(R\delta_i\) where
R is a free parameter, the learning rate. In this way, with each
iteration the weights become more optimal to producing a low
\(L\) and thus more accurate classifications. Assuming a network of
sufficient complexity to encode significant patterns and key features in
the data, this performance will generalise to examples outside of the
training set. If the dimensions of the network are not optimal, or the
training set is too small, overfitting can occur where low loss is
achieved on the training set but is not reflected in performance on
examples not seen by the network during training. Typically, training
examples are divided up into training, validation and test sets, where
the training set is used to train the network and update the weights, 
validation is used to measure progress during training and assist
in tuning parameters such as the learning rate, and the test set is
reserved for a final estimate of network accuracy using labelled examples
blinded from the network.

\section{Method}\label{sec:method}

Constructing a neural network-based system for lens-finding requires the
following steps. First, we assemble training sets. Due to the limited
number of known galaxy-galaxy lenses available, these consist of
simulated strong lenses and non-lens systems (see
section~\ref{sec:simulations}). We use the training set to iteratively
train two convolutional neural networks using the Keras Deep Learning
framework \citep{CholletKeras2015} on a GPU machine. We then take a
catalogue of 1.1 million sources selected to match the simulations in
\(g - i\) and \(g - r\) colour space and evaluate postage stamp images
of each galaxy with the neural networks, producing a score in the
interval (0, 1) for each image. We manually examine images with scores
greater than a chosen threshold and grade them 0-3, where 0 = not a
lens, 1 = possibly a lens, 2 = probably a lens, and 3 = definitely a
lens.

\subsection{Choosing the target source
population}\label{sec:catalog_colours}

Our science goal for the lens search is to assemble a population of
lenses with measurable Einstein radii at redshifts \(\gtrsim 0.8\) in
order to probe their total mass profiles in this redshift range.
Examining the spectral energy distribution (SED) of a typical lensing
galaxy, i.e.~a red, quiescent elliptical, 
we see that at redshift
\(\approx 0.8\) the rest frame UV dropoff is pushed almost entirely
redward out of the DECam \(g\)-band filter. Thus in this redshift range we
expect that a galaxy-galaxy lens with sufficient magnification to be
detectable would exhibit bright source flux in the g-band but would lack
a bright lens counterpart in the center of the image. This morphological
hint is something we hypothesize will be utilised by the CNNs (see
discussion in section~\ref{sec:discussion} below).

In this section we describe the method used to choose a subset of
sources in the Dark Energy Survey to search for lenses. We use catalogue
values to make these cuts, then test postage stamp images of selected
sources taken from DES Y3A1 coadd imaging. We restrict our search to a
subset of sources in the survey catalogue for two reasons. Firstly, it
reduces the amount of computational resources required, a significant
consideration for a survey with around \textasciitilde{}10TB of image
data. Secondly, even a hypothetical, extremely accurate lens finder
with a 0.1\% false positive rate would be expected to identify
300,000 false positives across a survey of this size, a number 2-3
orders of magnitude greater than the number of lenses we expect to
discover (see section~\ref{sec:completeness}). We therefore seek to
increase the purity of the sample by restricting the search to sources
we know are much more likely to be lenses than the average catalogued
galaxy.

\begin{figure*}
\centering
\includegraphics[width=0.90000\textwidth]{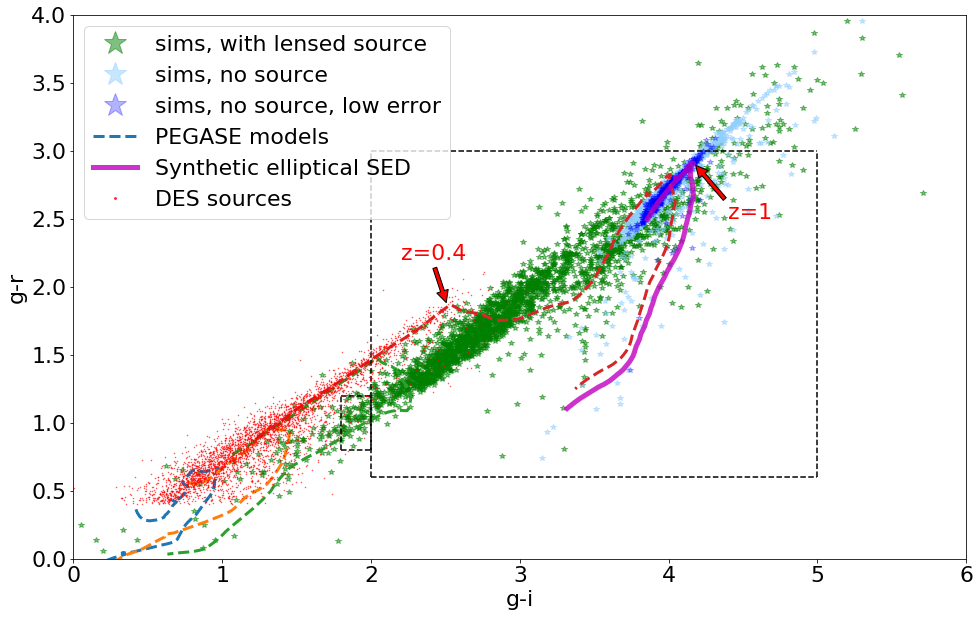}
\caption{The colours (in \(g-i\) and \(g-r\) ) of simulated lenses at
redshifts \textgreater{} 0.8, showing simulations without lensed source
(blue/cyan) or with (green). 
As the simulated ellipticals are faint or undetectable in g by design, there are large errors in 
the measured $g$-band magnitudes; this diagonal scatter (i.e. along the g axis)
is visible in the figure, compared to the raw colours of our \textsc{LensPop} 
synthetic 10 Gyr SED (redshifts 0.8-1.5 depicted in magenta).
Simulated lenses with photometric g-band magnitude errors \(< 0.2\) 
are depicted in dark blue, the rest in cyan. 
We depict a set of 
red through blue PEGASE.2 \citep{FiocPEGASEmetallicityconsistentspectral1999a}
template tracks, with progressively increasing amounts of recent 
star formation to illustrate where normal z<1.5 unlensed 
galaxies are expected to lie. The red dashed line is the PEGASE 10 Gyr
simple stellar population model (similar to our synthetic SED);
the arrows point out kinks in the colour track at redshifts 0.4 and 1.
A random selection of DES catalogue sources is depicted as red points, indicating
where the denser parts of the catalogue lie; our colour cuts are depicted as black boxes.
}\label{fig:colourcolour}

\end{figure*}

In catalogue space, ellipticals at these redshifts are very red and the
vast majority will lie at colours \(g-i > 3\) and \(g-r > 2\). This
serves as a starting point for our search for likely candidates.
However, the presence of a magnified lensed source, most commonly a
compact, blue, star-forming galaxy, will shift the system in colour
space to a degree difficult to predict from first principles given the
range of source and lens colours and magnifications we expect to see. In
order to constrain our catalogue search we use simulated lenses, the
production of which is detailed below in section~\ref{sec:simulations}.
We find that for a population of 10,000 simulated high-redshift elliptical
galaxies with simulated lensed sources superimposed, the distribution of
colours is as depicted in Figure~\ref{fig:colourcolour}. We depict the 
colours of our simulations with and without the lensed source.  
As the simulated
ellipticals are faint or undetectable in \(g\), there are large errors
in the measured \(g\)-band magnitudes; this scatter is visible in the figure,
compared to the raw colours of our synthetic 10gyr SED. 
Unlensed spirals are possible false positives. 

\begin{figure*}
\centering
\includegraphics[width=0.90000\textwidth]{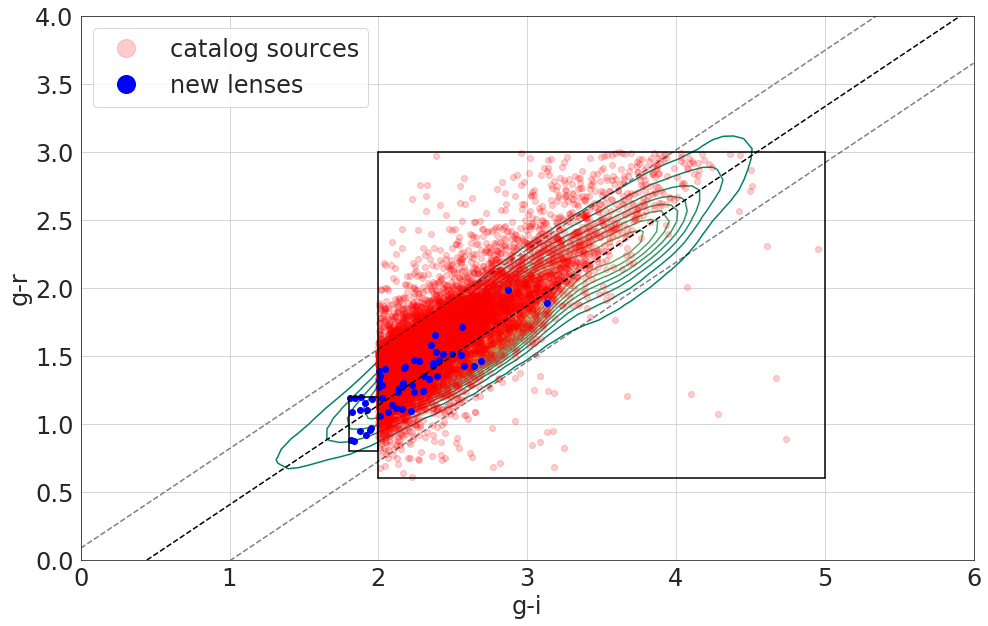}
\caption{Catalogue sources (green) and new lens
candidates (blue). Green contours indicated where simulated strong lenses lie. 
The dashed lines show a best fit colour-colour
relation for the simulated lenses, with 3-sigma lines shown. We choose
sources from the catalogue that are within the boxes depicted, where
\(2 < g - i < 5\), \(0.6 < g -r < 3\), and where \(1.8 < g - i < 2\),
\(0.8 < g -r < 1.2\).}\label{fig:colourcolour2}

\end{figure*}

The addition of
a lensed source shifts the simulated systems towards the blue end
of the spectrum by up to three magnitudes.
The colours used are the intrinsic colours of the simulated lens systems, 
with shot noise but without sky or any contaminants such as nearby objects.
Looking at the area of colour space where the majority of simulated
lenses lie, we build a catalogue as follows: We choose sources with
colours \(2 < g - i < 5\), \(0.6 < g -r < 3\), allowing for a 
large errors in measured \(g\)-band magnitudes for faint sources.
In order to test the
diminishing returns predicted by the simulations outside this region, we
supplement the catalogue with sources where \(1.8 < g - i < 2\),
\(0.8 < g -r < 1.2\), as depicted in Figure~\ref{fig:colourcolour2}. We
also restrict ourselves to sources where \(r_{mag} > 19\),
\(g_{mag} > 20\), \(i_{mag} > 18.2\) again following the distribution of
simulated lens luminosities. This represents less than 0.5\% of the
total survey catalogue. As we move bluer than this region of colour
space, the number of sources in the DES catalogue to examine increases
rapidly, and the number of simulated lenses decreases just as sharply.
We expect rapidly diminishing returns and so limit our search to this
region, which includes 93.4\% of the simulated lenses.
We discuss this further in section~\ref{sec:catalog_selection}.

We discard sources with undefined magnitude errors or flux errors in
\(gri\) bands, or where more than 400 pixels are masked out in the
100x100 postage stamps. We assemble a catalogue to search of 831,056 and
230,812 in the supplementary catalogue, for a total of 1,061,868 sources
selected from the complete DES catalogue.

\subsection{Generating simulations}\label{sec:simulations}

In order to optimise a neural network with millions of trainable
parameters (``weights'') we require a training set of sufficient size.
State-of-the art neural networks used in general computer vision applications 
require of order \(10^6\)
training examples for robust training \citep[e.g.][]{krizhevsky_imagenet_2012}. 
Given that the number of discovered lenses across all surveys and instruments is in the
hundreds, we must simulate lenses in order to create a training set of
sufficient size. We use a modified version of the \textsc{LensPop} code
described in Collett \citeyearpar{collett_population_2015} for this
purpose.

\textsc{LensPop} generates a population of synthetic galaxies with a
singular isothermal ellipsoid (SIE) mass profile and redshifts, masses
and ellipticities drawn from realistic distributions following the
\textsc{LensPop} methodology \citep{collett_population_2015}. Deflector
masses are drawn from the velocity dispersion function of SDSS
\citep{choiInternalCollectiveProperties2007a} without redshift
dependence and a constant comoving density out to redshift 2. Lens
colours assume a 10Gyr-old quiescent SED. Sources are elliptical
exponential disks with redshifts sizes and colours drawn from the COSMOS
sample \citep{ilbertCosmosPhotometricRedshifts2009}. Lens light is added
to the resulting image using the fundamental plane relation
\citep{Hydeluminositystellarmass2009} assuming a de Vaucolours profile
and the spectral energy distribution of an old, passive galaxy. We shift
the brightness profile of the sources by one magnitude brighter in all bands
to create a larger sample of detectable lenses. This makes the process
more efficient in terms of detectable lenses generated per second;
generating an unrealistically rich sample of bright, detectable lenses
is not problematic when our goal is simply to train our CNN and not
constrain lensing statistics in the real universe.

The \textsc{LensPop} code generates our synthetic population of lenses
and sources. The simulations are then pruned as follows. Firstly, lenses
with redshifts \(> 2\) and \(< 0.8\) are discarded. Lens images are then
simulated using \textsc{GRAVLENS}
\citep{KeetonComputationalMethodsGravitational2001} raytracing code.
Images in \(g\), \(r\), and \(i\) bands are produced with seeing
drawn from the DES Year 1 science verification data with a floor of
\(.9\arcsec\) in all bands; typical seeing of 1.1 - 1.2 \(\arcsec\).

\begin{figure*}
\centering
\includegraphics[width=0.95000\textwidth]{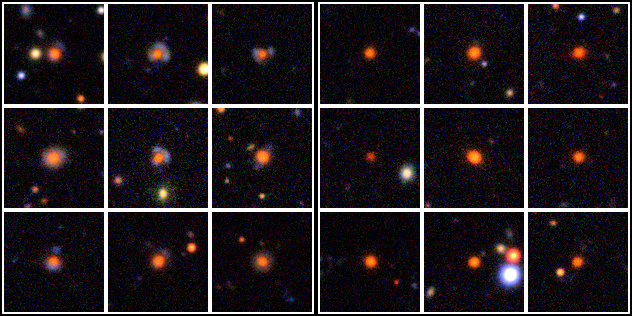}
\caption{Simulated lenses for the training set (RGB images from 
\(g\), \(r\), \(i\) bands). Left: With lensed
source. Right: Without lensed source (negative
examples).}\label{fig:tset}

\end{figure*}

\begin{figure}
\centering
\includegraphics[width=0.49000\textwidth]{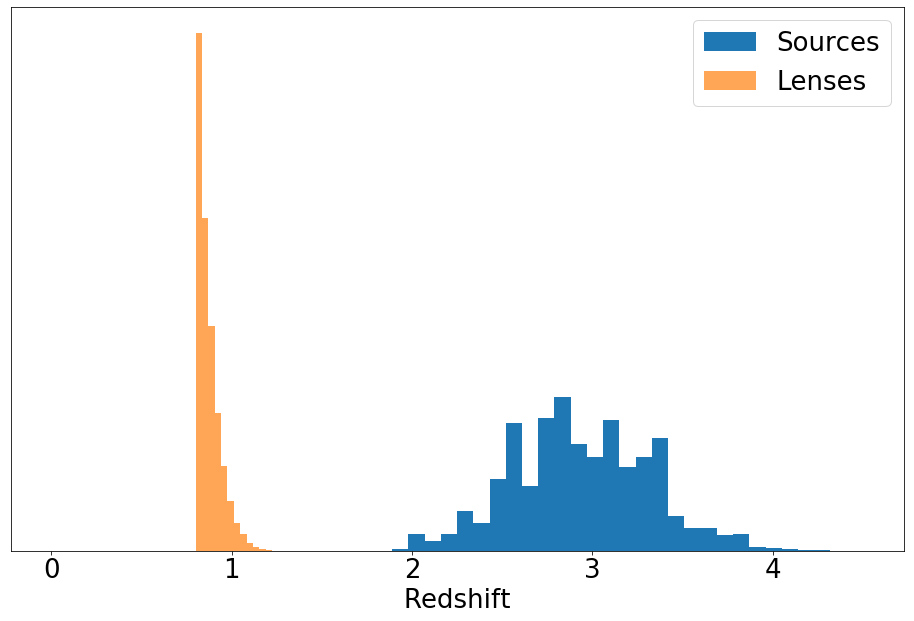}
\caption{Redshift distribution of lens galaxies and lensed sources for
simulations used in training the networks. This includes only lenses at
redshift > 0.8 and the associated sources.}\label{fig:nz}

\end{figure}

Simulated shot noise is added. Lenses with signal-to-noise \(< 3\),
Einstein radii \(<\) twice seeing and magnifications less than 3 are
discarded as they are unlikely to be detectable in DES imaging. We
generate two sets of images, as FITS files 100 pixels (\(30\arcsec\)) on
a side, the first with both the flux from the lensed source - positive
examples (``a strong lens'') and secondly, without - negative examples
(``no lensing depicted''). These simulated lenses are combined with
randomly chosen tiles from the DES imaging, to add sky and read noise,
stars, realistic background and foreground objects, artifacts, etc. We
assembled a training set of 250,000 images as depicted in
Figure~\ref{fig:tset}. A histogram of the redshifts of the simulations
is depicted in Figure~\ref{fig:nz}.

\subsection{Training Convolutional Neural Networks}\label{sec:training}

The convolutional neural networks were architected with four convolutional
layers with kernel sizes 11, 5, 3 and 3 respectively, and one fully
connected layer of 1024 neurons. The nonlinearity function is the
rectified linear unit (ReLU)\footnote{\(f(x) = max(x, 0)\)}; a dropout
\citep[see][]{HintonImprovingneuralnetworks2012} of 0.25 is added after
the last convolutional layer, and 0.5 between fully connected layers.
This network architecture is similar to industry standard network
architectures such as AlexNet \citep{krizhevsky_imagenet_2012}, but much
simpler than the most complex networks used for computer vision
(e.g.~ResNet \citep{HeDeepResidualLearning2016}, up to 1000 layers). The
network contains a total of 8,833,794 trainable weights. 
The number of
layers and their dimensions are free parameters, and an optimal
architecture is still a matter of some guesswork. This network
architecture was chosen based on previous experience
\citep{JacobsFindingstronglenses2017}, and was deemed fit-for-purpose
based on the high accuracy realised during the training process. A
deeper network could potentially result in higher training accuracy,
however the practical limitation appears to be the translation from
simulations to real sources see \ref{sec:false_positives}. A similar
network to the one presented here was used by the authors to enter the
Bologna Lens Finding Challenge
\citep{MetcalfStrongGravitationalLens2018} and placed third in the
detection of simulated lenses in multiband imaging.

\begin{figure*}
\centering
\includegraphics[width=0.85000\textwidth]{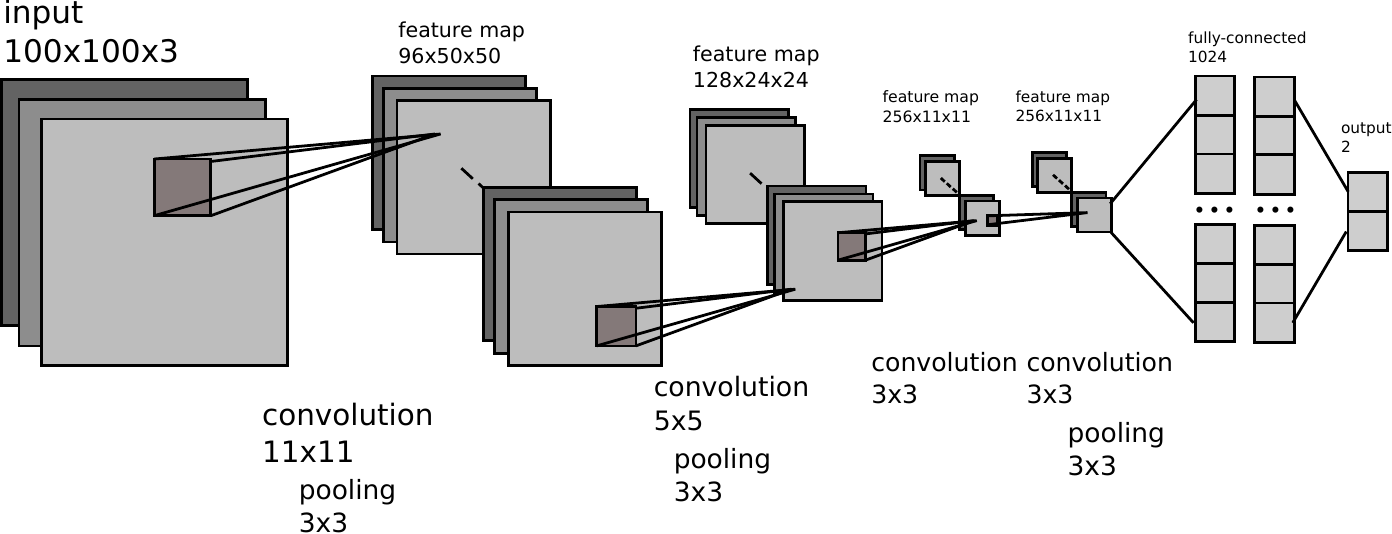}
\caption{Architecture of neural networks used: Four convolutional layers
with kernel sizes 11, 5, 3, and 3, and two fully-connected layers of
1024 neurons each.}\label{fig:network_arch}

\end{figure*}

The networks were implemented, trained and run using code employing the
Keras deep learning library and Theano numerical library
\citep{TheTheanoDevelopmentTeamTheanoPythonframework2016}.
Figure~\ref{fig:network_arch} depicts the network architecture; the description
of the Keras model is also included as an appendix. 

In total, 20 CNNs with these dimensions are trained, with differences
as outlined below. We create two
training sets, as summarised in Table~\ref{tbl:tsets}.
Training set 1 consists of 125,000 simulated lenses and
the same number of non-lensing elliptical galaxies. Training set 2
consists of 80,000 simulated lenses and 80,000 postage stamps of sources
chosen at random from our search catalogue
section~\ref{sec:search_catalog} as negative examples. With the first
training set, we ensure the network learns to reject simulations that do
not exhibit detectable strong lensing, forcing it to learn from the morphology
of lensing and not merely a characteristic of our simulations that inadvertently
distinguishes the simulations from real galaxies. With the second training set, the
networks will learn that objects we have not simulated - spirals,
mergers, stars, an so on - are to be considered non-lenses. Since we
expect only of order one lens in \(10^4\) sources, this negative
training set may be `contaminated' by a few actual lenses, but this will
not have a discernible impact on training since the contribution of each
training example to the weight updates is equal.

The use of two training sets with different non-lens images, as opposed
to a single larger training set combining both, has the advantage that
we can tune the weighting given to the contribution of the two training
sets when assembling a candidate set by choosing different score 
thresholds for the two networks. This gives more fine-grained control in
exploring the trade-off between purity and completeness and tuning the
size of the candidate set to examine.

\hypertarget{tbl:tsets}{}
\begin{table*}
\centering

\caption{\label{tbl:tsets}Training sets used to train neural networks.}

\begin{tabular}{@{}llll@{}}
\toprule

Training set & Pos.~examples & Neg.~examples & Size \\\midrule

TS1 & simulations & simulations & 250,000 \\
TS2 & simulations & real galaxies & 160,000 \\

\bottomrule
\end{tabular}

\end{table*}

For each of these two training sets, we divide each into 10
equally-sized subsets (folds). For each fold, we train a network
reserving that fold of the data as a validation set - not used for
training, but used to measure training progress - and the remainder as
the training examples. We thus obtain 10 networks trained on different
subsets of the training examples to hand. This process is known as
k-fold cross-validation \citep[see][ for a detailed
description]{RefaeilzadehCrossValidation2009}. There is some
stochasticity in the training process; the initial weights are
randomised, the order in which the training set is fed to the network is
also random, and by using slightly different training sets, each network
thus trained will score candidates slightly differently. Using an
ensemble allows us to smooth out the effects of outlier scores; we use
the mean score from the 10 trained networks in selecting candidates.
More than 10 networks per ensemble are unlikely to add additional
information, but require GPU time to train. It has been shown
\citep{HansenNeuralnetworkensembles1990, KroghNeuralnetworkensembles1995}
that using an ensemble of neural networks in this way 
can provide a significant
boost to the accuracy of the system, e.g.~a 2\% increase in
classification accuracy over the best performing network by an ensemble
\citep{JuRelativePerformanceEnsemble2017} - particularly if the networks
are trained with different training data \citep{GiacintoDesigneffectiveneural2001}.

The networks are trained on FITS data in three bands (\(g, r, i\)),
passed to the networks as 32-bit floating point values. The FITS data,
which is background-subtracted, is further
normalised so that across the training set, the mean value is zero and
99.7\% of the values lie between -2.5 and 2.5\footnote{\(X' = (X - \mu)/\sigma\)}.
This is shown to optimise convergence by the training algorithm
\citep{LeCunEfficientBackProp1998}.

We train the networks until further iterations no longer decrease the
loss value on the validation set. At each epoch (iteration through the
training set), we test the accuracy of the network on the training and
validation sets, and calculate the loss for each. We halt training when
the loss on the validation set has decreased by less than a parameter
\(\epsilon = 0.0001\) for six epochs. Further training beyond this point
is likely to lead to over-fitting to the training set.

\subsection{Scoring and sorting candidate sources}\label{sec:scoring}

Our target data set for the lens search is Dark Energy Survey 
\citep[DES;][]{diehl_dark_2014,flaugher_dark_2015,diehl_dark_2016}
Year 3
coadd images \citep{AbbottDarkEnergySurvey2018,MorgansonDarkEnergySurvey2018}. 
This imaging consists of 10,346 tiles over 5,000 square
degrees of sky. The number of epochs is \textasciitilde{}4-6 per coadd
object per band, with a limiting magnitude in \(r\) of 24.9 and a pixel
scale of \(0.263\arcsec\)/pixel. The mean seeing is \(1.06\arcsec\) in
\(g\) \citep{Diehl018jtu}. We generate postage stamps in \(g\), \(r\), \(i\) and bands of
dimensions 100x100 pixels for each of the million sources in our target
catalogue. Each of the postage stamps is scored using the pre-trained
CNNs, to produce two scores in the interval (0, 1) corresponding to the two
different training sets. We then examine the
distribution of scores, and choose thresholds for each score to produce
a subset of our catalogue for visual examination by human experts. We
choose the threshold such that the candidate set is of a size that can
be examined in a few hours, i.e.~a few thousand images. RGB images of
each source are examined by eye (by authors CJ, KG and TC) and graded
using software, LensRater, developed for this purpose\footnote{https://github.com/coljac/lensrater}.
We rate the candidates as 0) unlikely to contain a lens, 1) possibly
containing a lens, 2) probably containing a lens and 3) almost certainly
containing a lens. We then take the mean grade and assemble our final
candidate catalogue from those graded 2 and above. In this paper we
define false positives as any candidates that we judge to be below grade
1. We then estimate the completeness of our sample of lens candidates.

\subsection{Estimating photometric
redshifts}\label{estimating-photometric-redshifts}

The objects we discover in our search are lens candidates. In the
absence of spectroscopic follow up, we cannot know how many of them are
genuine strong lenses, and of those that are, how many are in our target
redshift range. In order to make a first-order approximation regarding
the second question, we calculate photometric redshifts of the lens
galaxies. We use the \textbf{BPZ} (Bayesian Photometric Redshifts)
photo-z package\footnote{http://www.stsci.edu/\textasciitilde{}dcoe/BPZ/}.
As inputs to the photo-z code we use colours measured from the DES Y3
coadd images in \(griz\) with apertures fit manually to the galaxies
(excluding blue source flux), with mag errors taken from the DES
catalogue. We quote the best-fit and \(2\sigma\) uncertainties output by
\textbf{BPZ}.

\subsection{Estimating the completeness of the
sample}\label{sec:blinded}

Our workflow involves the evaluation of machine-selected candidates
by human astronomers for follow-up. The optimal sample would therefore
include all sources that a human astronomer would grade as probable
or definite lenses, and not those that would be graded otherwise, whether
or not they are, in reality, strong lenses. The completeness of our sample,
as a measure of what can realistically be detected in the imaging we are
searching, is a function of what an astronomer can discern with confidence
from a composite RGB image used for evaluation.

Collett \citeyearpar{collett_population_2015} used simulations to estimate
the number of strong lenses discoverable in DES coadd imaging. 
Simulating the survey sky, using detectabity criteria of 
signal-to-noise in \(g\) greater than 20, magnification greater than 3,
and an Einstein 
radius greater than the seeing (\textasciitilde{}\(1\arcsec\)), Collett
predicts \textasciitilde{}1300 lenses should be discoverable by inspection
of the images.
These detectable lenses had a mean lens redshift of 0.42; 8\% (\textasciitilde{}110)
were at redshift 0.8 to 2.

How many of these theoretically-detectable lenses would actually be selected as good
candidates by a human astronomer following our lens-finding pipeline is a
testable question. To better understand this threshold, we collect one further 
piece of data.  We assembled a set of 5000 postage stamps containing
2500 real galaxies, 1000 simulated lenses, 1000 simulated ellipticals
and 500 simulated ellipticals with unlensed blue sources nearby 
(``phonies'') and presented these,
blinded, to authors TC and KG to evaluate. We then examine the number of
simulated high-redshift lenses graded highly by the inspectors.
Measuring the fraction of simulated lenses that were rated highly 
assists us in making an estimate of the true number of lenses we can expect to find in the
survey using our automated pipeline. Of high-redshift simulated lenses examined, 51\%
were given grade 0, indicating that estimates of
detectability are highly dependent on image quality and grading methodology, 
and can easily be overestimated.
We discuss this further in section~\ref{sec:completeness}.

\section{Results}\label{sec:results}

\subsection{Training neural networks}\label{training-neural-networks}

\begin{figure}
\centering
\includegraphics[width=0.49000\textwidth]{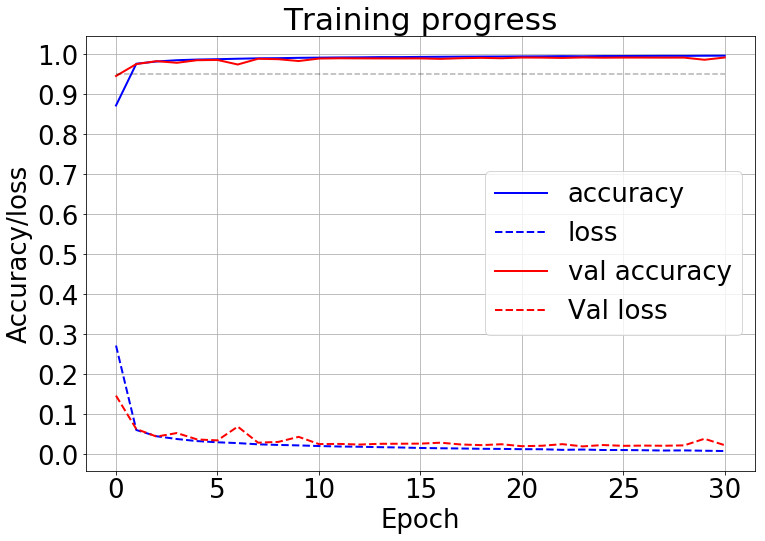}
\caption{Training of a neural network, demonstrating convergence on high
accuracy and low loss. The dashed lines show the value of loss function,
evaluated over the image set, and the solid line the classification
accuracy. Blue: Loss/accuracy on the training set. Red: Loss/accuracy on
validation set of images not used for
training. The curves for other networks are similar and so are not shown.}\label{fig:training_progress}
\end{figure}

Two ensembles of neural networks were trained as described in
section~\ref{sec:training}. For the first ensemble, trained on simulated
lenses with and without lensed sources, the training converged after
\(30 \pm 1\) epochs in each case. The accuracy (the fraction of a sample classified
correctly: true positives + true negatives a divided by number of items 
tested) on the respective validation
sets of the 10 networks in the ensemble was \(98.6 \pm 0.1\%\). The
training progress for a single network is depicted in
Figure~\ref{fig:training_progress}; after a single epoch, the training
accuracy was 87\%, converging slowly on the final value. On the second
training set, composed of simulated lenses and random sources from the
catalogue, training converged in fewer epochs, \(20 \pm 2\), with a
validation accuracy of \(99.4 \pm 0.1\%\).

\begin{figure}
\centering
\includegraphics[width=0.49000\textwidth]{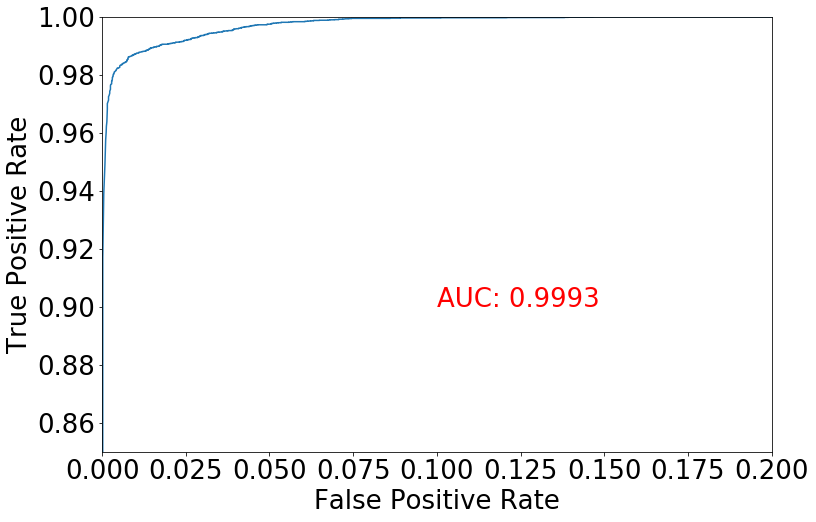}
\caption{Receiver Operating Characteristic curve for the CNN trained on training set 1, 
consisting of simulated strong lenses and simulated elliptical galaxies without visible lensing. This curve shows the trade off between a desired true positive rate and the number of false positives produced by the network for different values of the score threshold. 
The area under the curve (AUC) for this network is 0.9993. For training set 2, the AUC is 
0.9998, so the curve is not shown.}
\label{fig:roc_curve}
\end{figure}

In Figure~\ref{fig:roc_curve} we depict the Receiver Operating Characteristic curve for the 
first network, trained on simulated lenses and non-lenses, when evaluated on examples
not used during training. This figure depicts the trade-off between the true positive
rate and the false positive rate achieved for different values of the score threshold.
A perfect system would include the point at (0, 1), namely zero false positives and
all true positives, and have an area under the curve (AUC) of 1. 
The AUC is for the first network is 0.9993; for the second, it is 0.9998, and so the 
curve is not shown.

The total training time was approximately 40 hours for the first
ensemble and 24 hours for the second, trained on an NVidia K80 GPU and
Intel Xeon E5-2698 cpu with 12GB RAM and a batch size of 128 images.

\subsection{Scoring catalogue sources and selecting a candidate
set}\label{sec:search_catalog}

Scoring a batch of 128 100x100 pixel FITS images in three bands took
\textasciitilde{}3ms. With the overheads of loading the files into
memory, and scoring with 20 networks, scoring the 1 million sources in
our catalogue took approximately six hours. The 254GB of images were
stored in HDF5 databases in 15GB chunks and the CNNs were able to load
the images in batches using the HDF5 files directly, a faster process
than working with 1 million or more individual files.

\begin{figure}
\centering
\includegraphics[width=0.45000\textwidth]{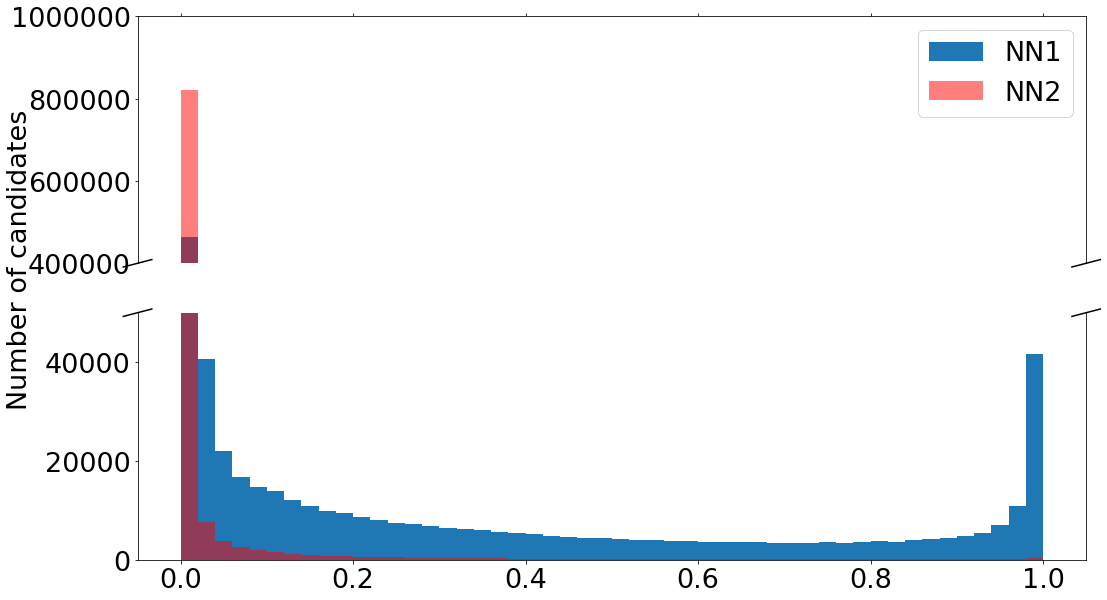}
\caption{Distribution of scores of sources scored by CNNs. For the
first ensemble, of 1.1 million sources, 3144 had a score of 1.0
(definitely a lens), 668408 had a score \textless{} 0.01. For the
second, there were no perfect 1.0 scores, and 810,604 scored \textless{}
0.01. There are 358 sources in the final bin with score \textgreater{} 0.98.}\label{fig:allscores}
\end{figure}

We scored each of the 1.1 million postage stamps with all of the 10
trained networks in each of the two ensembles. We took the mean score
from each ensemble to produce two scores for each image. Of the
1,061,868 sources scored by the first ensemble of networks, 576,025
(54\%) were scored less than 0.01; and by the second ensemble 967,348
(91\%). The first ensemble scored 35,332 sources above 0.99; the second,
only 433. The scores are summarised in Table~\ref{tbl:score_summary} and
a histogram depicting the distribution of scores is presented in
Figure~\ref{fig:allscores}.

\hypertarget{tbl:score_summary}{}
\begin{table*}
\centering

\caption{\label{tbl:score_summary}Distribution of CNN scores for the
two ensembles. }

\begin{tabular}{@{}lll@{}}
\toprule

Score & Ensemble 1 & Ensemble 2 \\\midrule

\textless{} 0.01 & 576,025 & 967,348 \\
\textgreater{} 0.5 & 156,776 & 9328 \\
\textgreater{} 0.99 & 35,332 & 433 \\
\textgreater{} 0.999 & 10,847 & 97 \\

\bottomrule
\end{tabular}

\end{table*}

\begin{figure}
\centering
\includegraphics[width=0.45000\textwidth]{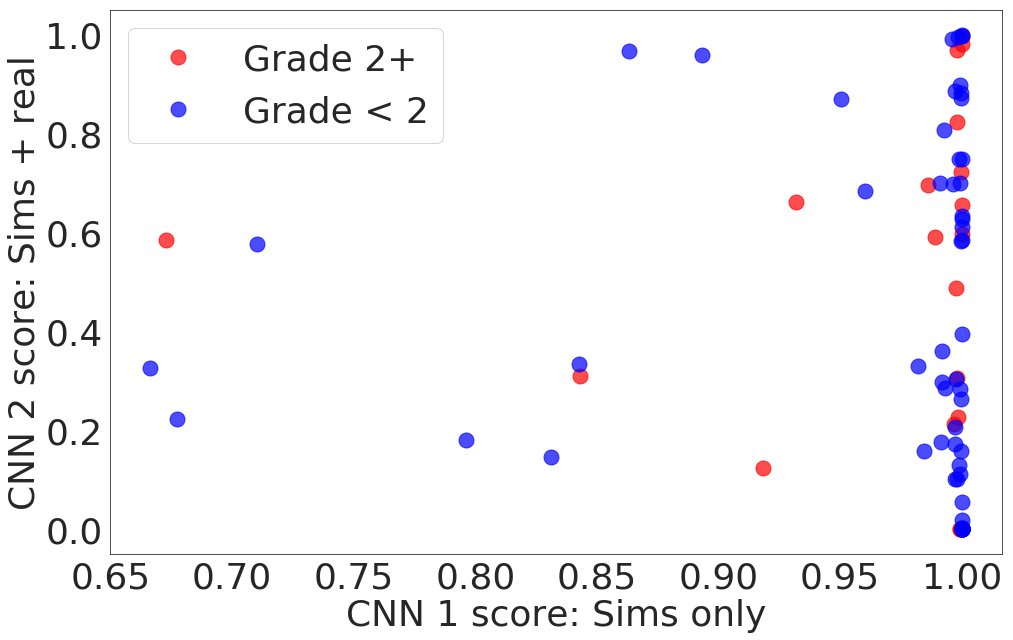}
\caption{Scores received from the CNNs by the most highly-graded 
lens candidates. Red: Graded 2-3 ("probably" or "definitely" lenses). 
Blue: Scores below 2 (possibly" lenses). Some of the best candidates were
scored as low as 0.2 by the network trained on simulations and real
galaxies.
}\label{fig:score_plot}
\end{figure}

Due to the subtleties of lensing morphology in this redshift range, and
the large number of sources evaluated, false positives are a concern. We
wish to produce a candidate set for visual inspection that is as small
(pure) as possible while containing the majority of the detectable
strong lenses in the survey (high completeness). As the discoverable lenses
are not known \emph{a priori}, evaluating completeness is only possible
in approximation and after evaluation by eye has been completed (see
section~\ref{sec:discussion} below).

We choose candidates for visual inspection by selecting score thresholds
and examining candidates that scored higher than this number by the
networks. The thresholds \(t_1\) and \(t_2\) are free parameters; the
scores \(s_1\) and \(s_2\) are output by the two CNN networks for each
source tested. We examine candidates where, for that source,
\(s_1 > t_1\) and \(s_2 > t_2\). This filters many sources scored highly
by one network but not the other.

We examine candidate sets as per Table~\ref{tbl:search_summary}. With
thresholds (0.65, 0.1) we obtain 3,582 images to examine; with threshold
(.9999, 0) a further 1,841 candidates; and in the area of the extended
catalogue, 1,878 images with thresholds (0.95, 0.55) for a total of
7,301 images. We choose these candidate sets so as to explore the
relative contribution of the two CNN ensembles while returning a
manageable number of candidates. Following inspection of these
candidates, author CJ examines a further 9,428 candidates with scores
above thresholds (0.999, 0) for a total of 16,729 images. The set with
scores (0, .999) contained only 49 images, all false positives.

\subsection{Examining candidate
lenses}\label{examining-candidate-lenses}

Of 16,729 the candidates examined, 250 had a grade \textgreater{} 0, 87
\(\geq 1\) and \(29 \geq 2\). With grade 0 candidates, we have an
overall false positive rate (false positives = highly scored non-lenses)
of 98.5\% amongst the candidates we
reviewed. Of the candidate sets we reviewed, the purest was the 3,582
candidates with scores 
\(s_1 > 0.65\) and \(s_2 > 0.1\), which yielded 43
candidates with grades \textgreater{} 1 for all examiners. Overall time
taken to examine candidates is approximately five hours of astronomer
time. Of the 87 candidates identified with scores \(\geq 1\), 4 are
known from a previous search \citep{DiehlBrightArcsSurvey2017}.

\begin{figure*}
\centering
\includegraphics[width=0.85000\textwidth]{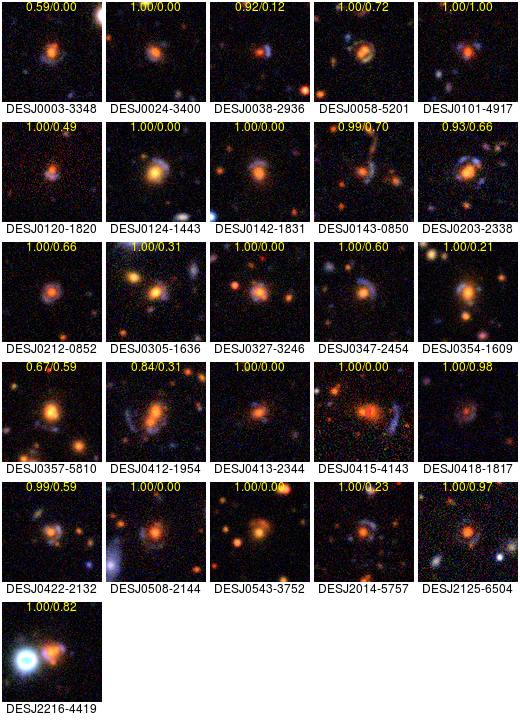}
\caption{26 new lens candidates, with a grade \(\geq 2\), discovered in
DES imaging data using CNNs. The scores from the two CNNs are shown in yellow
text.  The candidates are described in
Table~\ref{tbl:new_candidates}.}\label{fig:candidates_1}

\end{figure*}

The lenses with a grade \(\geq 2\) are presented in
Figure~\ref{fig:candidates_1} and those with \(1 \geq \text{grade} < 2\)
are shown in Figure~\ref{fig:candidates_2}. The candidates are
summarised in Table~\ref{tbl:new_candidates}, including with the
photometric redshifts for the lenses with \(2\sigma\) errors estimated
by \textbf{BPZ}.

The scores the candidate lenses received from the networks are presented
in Figure~\ref{fig:score_plot}. Most candidates received scores of approximately
1.0 from the CNN trained on simulations, but were more evenly distributed
in their scores from the second CNN trained on simulations and real galaxies.
There is no significant difference in CNN scores by grade of lens candidate.
The mean scores for candidates of grade 2+ were .97 and .39 for 
the two networks; for grade \(< 2\), the mean scores were .87 and .42
respectively.

\hypertarget{tbl:search_summary}{}
\begin{table*}
\centering

\caption{\label{tbl:search_summary}Summary of candidate sets examined.
These candidate sets were selected by the neural networks with scores
greater than the thresholds \(t_1\) and \(t_2\) and examined by the
authors. New candidates indicates candidates that are unique to that
search; search 4 contained 22 candidates, but only four that were not
found in the other searches, demonstrating the rapidly diminishing
returns. Searches 1, 2, and 4 were applied to the larger source catalogue,
search 3 was applied to the extended catalogue only, as described in
section section~\ref{sec:catalog_colours}}.

\begin{tabular}{@{}lllllll@{}}
\toprule

Search & size & \(t_1\) & \(t_2\) & candidates \textgreater{}= 2 & New
candidates & Purity \\\midrule

Search 1 & 3582 & 0.65 & 0.1 & 11 & 43 & 1.2\% \\
Search 2 & 1841 & 0.9999 & 0.0 & 5 & 15 & 0.8\% \\
Search 3 & 1878 & 0.95 & 0.55 & 6 & 21 & 1.1\% \\
Search 4 & 9428 & 0.999 & 0.0 & 3 & 4 & 0.04\% \\

\bottomrule
\end{tabular}

\end{table*}

Our catalogue was selected by examining the combined lens and source
colours of simulated lenses (section~\ref{sec:catalog_colours}).
Figure~\ref{fig:colourcolour2} depicts the position in g - r and g - i
colour space of the new lens candidates, as well as the simulations and
sources from our search catalogue.

We include one candidate, DESJ0003-3348, discovered serendipitously in
the control sample inspected in section~\ref{sec:blinded}. It received
scores of .59 and 0.00 from the two networks respectively.

\section{Discussion}\label{sec:discussion}

\subsection{Efficiency of the method}\label{efficiency-of-the-method}

Convolutional Neural Networks have proven themselves in a variety of
computer vision problems both broadly and within astronomy, including in
other lens finding applications. Here we also find that they performed
well on a more targeted lens search, producing dozens of high quality
candidates with a few hours of astronomer inspection time. Inspecting
the lenses with LensRater (section~\ref{sec:scoring}), we find that
examining 3000 candidate images per hour is a sustainable rate. We
examined approximately 7300 postage stamps, or 2.5 hours, in selecting
the catalogue of new lens candidates presented in
section~\ref{sec:results}. Thus, assuming all candidates are genuine
lenses, we discover \textasciitilde{}30 genuine lenses in our redshift
range per hour of astronomer time and achieve completeness close to
100\% (see below) in a few hours. In comparison, examining the entire
catalogue of 1 million lenses would take over 13 days at this rate, and
11 years for all sources in the survey.

We can almost certainly increase the completeness of our catalogue by
examining more potential candidates. However, as we reduce the neural
network score threshold to examine, the size of the candidate set
increases exponentially, as does the time investment required for each
additional candidate. As the candidates become less obvious to the human
eye (fainter, arc-like features more subtle), so does the number of
false positives increase. We examined 7,301 candidates and identified 83
probable or definite lenses (four of which are previously known).
Examining a further 9,428 candidates uncovered only four more credible
lenses in addition to those already identified. We conclude that these
diminishing returns indicate our sample is relatively complete at this
point.

\subsection{Catalogue selection}\label{sec:catalog_selection}

We restrict our search to postage stamps of a subset of sources in the
DES survey catalogue. The catalogue cuts, in \(g-i\) and \(g-r\) colour
space (section~\ref{sec:catalog_colours}), were chosen by reference to
the integrated colours of our simulated strong lenses in the desired
redshift range. Figure~\ref{fig:colourcolour2} depicts the locations of
both the simulations and the new lens candidates in this space.
We find
good agreement between the new candidates and the colours predicted by
the simulations. The candidates we present are significantly closer to
the area in the space where the simulations reside as opposed to the
catalogue sources more generally which exhibit greater scatter.
Searching a smaller catalogue that conformed more closely to the
contours of the simulated lenses would therefore seem like a promising
avenue to yield a purer sample without sacrificing completeness. Of our
total catalogue of 1.1 million sources, 36\% lie \(> 3\sigma\) from the
line best fit to the lenses. Although the CNNs perform some of this
pruning for us (of our candidate sets examined, 85\% lie within this
region), some time saving is achievable here. Of the candidates in our
catalogue, all lie within the \(3\sigma\) limit.

The density of lens candidates increases towards the bluer end of the
cuts we made in both colour dimensions. This suggests widening our
search may yield further candidates. However, this area of the colour
space contains a much greater density of catalogue sources; for
instance, while a million sources exist in the range we chose, an
additional 2.5 million sources are present if we go .5 mag bluer, and 10
million sources at 1 mag bluer in each axis. Thus, assuming a constant
rate for false positives, we would expect our purity to drop by a factor
of 2.5 - 10, yielding rapidly diminishing returns. The density of
simulated lenses was lower in this part of colour space; 91\% of
simulations are in the original catalogue area, and 5\% in the
supplementary catalogue. Blueward of the catalogues we searched, many
spiral galaxies are to be found, and we expect that a higher false
positive rate would accelerate the diminishing returns in extending the
search in this direction.

\begin{figure*}
\centering
\includegraphics[width=0.85000\textwidth]{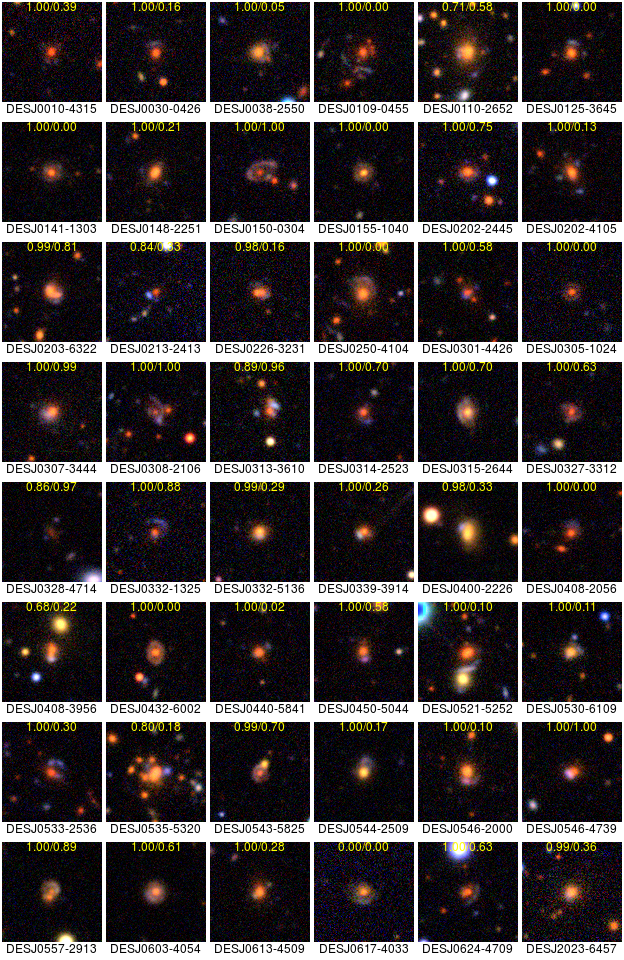}
\caption{}\label{fig:candidates_2}

\end{figure*}

\begin{figure*}
\centering
\includegraphics[width=0.85000\textwidth]{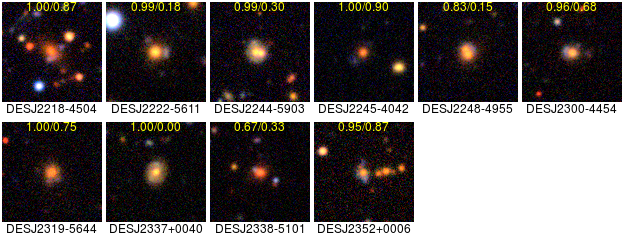}
\caption{58 new lens candidates, with a grade \(1 \geq \text{grade} > 2\),
discovered in DES imaging data using CNNs. The CNN scores are shown in yellow text.
The candidates are described
in Table~\ref{tbl:new_candidates}.}\label{fig:candidates_2}

\end{figure*}

\subsection{Completeness of the candidate sample}\label{sec:completeness}
In Jacobs et al \citeyearpar{JacobsFindingstronglenses2017}
we used CNNs to search CFHTLS, which had been the subject
of several fruitful lens searches previously, including visual
inspection of the entire survey area (\(171 \text{deg}^2\)) by citizen
scientists. Using a catalogue of lenses discovered in previous searches,
we were able to estimate a completeness of 21-28\% in a candidate set of
2465 sources.  Estimating completeness of our current sample is more 
difficult as we have no pre-existing
reference sample of high-redshift lenses in the survey footprint against which
to compare, and must rely on simulations to estimate the number of discoverable lenses.

The number of detectable lenses in a survey is a function of both 
the depth and seeing of the imaging, and the methodology used to examine
sources. Collett \citeyearpar{collett_population_2015} modelled the number of
detectable lenses in DES, and
estimated up to 1300 could be found, of which 110 are in our target redshift range.
This estimate assumes an optimal stacking strategy, and inspection of lens-subtracted
images, which we were unable to perform due to difficulties in modelling the PSF of
the coadd imaging.

We used LensRater, the candidate ranking pipeline 
described in section~\ref{sec:blinded}, to evaluate a mixture of simulated
lenses, potentially confusing chance alignments, and real galaxies.
For simulations at all redshifts, only 20\% received a grade \textgreater{} 0 by a human inspector.  Of 500 simulated high-redshift lenses in the sample with Einstein radii \textgreater{} ~\(2\arcsec\), 
247 (49\%), 99 (19.8\%) and 24 (4.8\%) received grades 1, 2, 3 respectively.
Detectability is aided at higher redshifts as the effective radius and apparent luminosity of the lens are smaller, allowing for greater image separation between source and lens.  The estimate of 130 lenses from \citep{collett_population_2015} includes lenses with smaller Einstein radii (59\% are smaller than \(1.25\arcsec\)), so this gives us an upper bound on the number of lenses we expect to find.
247 (49\%) received a grade \(\geq 0\), 99 (19.8\%) received a grade \(> 1\), and 24 (4.8\%) received a grade of \(\geq 2\).  
Detectability is aided at higher redshifts as the effective radius and apparent luminosity of the lens are smaller, allowing for greater image separation between source and lens.  The estimate of 130 lenses from \citep{collett_population_2015} includes lenses with smaller Einstein radii (59\% are smaller than \(1.25\arcsec\)), so this gives us an upper bound on the number of lenses we expect to find.
We therefore conclude that in DES, of order a few tens of high-redshift lenses
have the signal-to-noise and image separation required to be selected
confidently by a human inspector.
Our test on simulations can also give us some indication of the 
reliability of inspection grades.  Of those with grade 3, 100\% 
were simulated lenses; for grade 2, 98\%; and
for grade 1, 93\%. Conversely, only 2\% of the "phonies" received a grade
\textgreater{} 0.

To estimate the completeness of our search we also need to know
how many of the discovered lenses lie in the targeted redshift range.
The reliability of the photometric redshifts is limited by the number of
bands, contamination with flux from the blue lensed sources, and the
Bayesian priors used by the code. Due to the comparative rarity of massive
galaxies, the prior on the redshift distribution in BPZ strongly
penalizes elliptical galaxies with i\textasciitilde{}18 being beyond
z\textgreater{} \textasciitilde{}0.8. However, since our galaxies are
selected as strong lenses they must be massive and likely live in the
bright tail of the luminosity function. We therefore expect that the BPZ
prior is biasing the photometric redshifts low, but we have not quantified this
effect.  Of our sample of 84 candidates, 76 are within the 
targeted redshift range within
the quoted \(2\sigma\) errors, and 28 (33\%) within \(1\sigma\). From
this, we conclude that a sizeable fraction of our candidate set
are within the right redshift range, independently of whether they are
in fact strong lenses. 

Of the lens candidates presented in 
\citet[][]{DiehlBrightArcsSurvey2017},
a previous search of the DES imaging (see Section~\ref{comparison-to-other-des-strong-lens-searches}), 102 fall within out catalog. Of these, 33 
are galaxy-scale lenses consistent with our grading scheme but were
not detected in our CNN-based search. This may indicate that our 
completeness estimate is high. Another possible explanation is
that the CNNs have correctly filtered for redshift. None of 
these 33 candidates have published photometric redshifts 
greater than 0.8, and only two have redshifts less than
two times the stated redshift error below 0.8.

Here we present a catalogue of 84 candidate lenses; 26 have a grade
\(\geq 2\). Based on the above, we expect that the majority
of our sample will be confirmed as lenses, but spectroscopic followup 
will be required to constrain the fraction that are in the correct
redshift range. Based on the photometric redshifts, we expect a at 
least few tens of candidates to be confirmed,
which is of a similar order to the expected number of discoverable lenses.
Although it is not possible to constrain the error on this
estimate until follow-up is undertaken, the search seems likely to
increase the number of known lenses at high redshift by a factor of a
few.

If this result is confirmed, this also represents an
improvement in purity and completeness over previous searches. Although
there is still room to improve the method, we attribute the improved
performance (compared to \citet{JacobsFindingstronglenses2017})
to the use of CNN ensembles, improved training set
simulations, and the targeted search, which constrains the morphological
variety of the lenses sought, particularly in lens colour.

We note our candidates include one, DESJ0543-3752, with a red arc. This
indicates that while the CNNs clearly make use of both colour and
morphology, a clear signal in only one can still produce a high score.

\subsection{False positives}\label{sec:false_positives}

After testing trained networks on simulated lens images, the neural
networks are able to distinguish lenses from non-lenses with high
accuracy. Selecting images with scores greater than 0.5 as lenses, the
trained networks have accuracy between 98.6 and 99.4\% (for networks 1
and 2 respectively). If this performance translated perfectly to the
real survey imaging, we would expect that for 1 million sources
examined, we would achieve a completeness of \textasciitilde{}99\% of
the lenses in our catalogue - approximately 100 - and 10,000 false
positives (a purity of 1\%). Setting aside candidates that could be
lenses but are of low quality (score 1), we examined 7,301 sources to
find 52 lens candidates, a purity of 0.7\%. By that measure, the CNN
search results roughly reflect the performance expected from training.
The majority of the sources in this candidate set can be immediately
rejected by human astronomers. This implies a significant reduction in
false positives ought to be possible. Since real-world performance now
approximates the training performance, we conclude that investigating
the use of deeper and more complex networks, as well as improving the
simulations, may be warranted.

The false positives in the sample, i.e.~sources we rate as very unlikely
to be a lens/having no discernible features of strong lensing, exhibit a
wide variety of morphologies, but we can identify a few clear trends:
\begin{itemize}
    \item \emph{Blue near red}: sources of plausible colours, but no obvious
        morphology that would suggest strong lensing (\textasciitilde{}10\%);
    \item \emph{Low signal to noise}: Faint sources with apparent blue flux but insufficient
information present to clearly indicate lensing (\textasciitilde{}25\%); 
    \item \emph{Imposters}: Blue
spiral arms and other features that mimic lensing arcs (\textasciitilde{}5\%); 
    \item \emph{Unclear}: Some irregular sources don't resemble typical examples from either
category, and so the CNNs' best guesses are undefined (\textasciitilde{}60\%).
\end{itemize}

\begin{figure*}
\centering
\includegraphics[width=0.70000\textwidth]{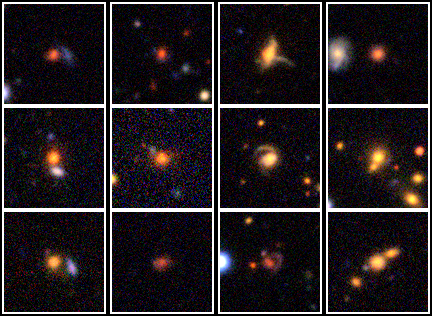}
\caption{False positives amongst candidate lenses. Left: Blue near red
objects. Second from left: Low signal to noise objects. Second from
right: False arcs. Right: Objects with no clear confusing
feature.}\label{fig:false_positives}

\end{figure*}

A representative sample is depicted in Figure~\ref{fig:false_positives}.
In searches aimed at finding lenses at other redshift ranges, we find
that spiral and ring galaxies form a large fraction of false positives,
as (for instance) blue star-forming regions in the arcs of spiral arms
can trigger the arc-detection features of the neural network strongly.
In this search, although spirals are present in the false positives,
they form a smaller fraction of the false positives we examined. Given
the colours and morphology of lenses at the higher redshift range, we
expect fewer spirals - morphological similarities notwithstanding - will
activate the networks strongly enough to achieve a high probability
score.

The false positives suggest two deficiencies in the training set.
Firstly, there may be too many simulated lenses that while theoretically
detectable, would not be graded highly on inspection by a human expert.
Since such lenses, when detected in the survey imaging, make poor
candidates for follow-up, we may wish to train networks instead to
reject them. Secondly, we are training the networks to place all
candidates in one of only two categories, lens or non-lens. Highly
irregular objects, which do not resemble typical examples of either
lensing or non-lensing objects, receive unpredictable scores. Despite
their rarity, a future training set could include a greater proportion
of irregular galaxies; however, by their nature it is uncertain how
successful a CNN would be at learning features from these objects.
Training the networks to place objects in more than two categories may
improve the situation.

Internally, the neural networks create a highly non-linear decision
boundary in the parameter space of all possible images, in this case
30,000 dimensions (100x100 pixels in three bands). Nguyen et al
\citeyearpar{NguyenDeepneuralnetworks2015} demonstrated that in
traditional computer vision applications using deep neural networks, it
is possible to construct images that appear to be white noise to a human
observer but strongly activate the networks for a particular image
category. This implies that if we examine enough noisy images, as we
will with large surveys, we will encounter some which, despite their
appearance to a human being, contain a configuration of values that
activate a part of the network strongly indicative of one of the two or
more defined categories. To enhance the purity of lensing searches in
future surveys, we seek false positive rates of order 1 in 100,000 or
better - an ongoing challenge when noisy images may activate by chance
particular parts of a trained network that indicate lensing. Further use
of ensembles of networks may mitigate this problem.

In the preceding discussion, we have considered false positives to be
candidates that a human inspector deems unlikely to be a strong lens.
However, some of these false positives are likely to be strong lenses, only
of a sort a human inspector would not grade highly. It is possible that
with improved inspection tools to aid the inspector, such as lens-subtracted
images, a human would be better able to identify lenses that the networks
score highly but are difficult to spot in the RGB images of the sort we use here. 

\subsection{Choosing a candidate set}\label{choosing-a-candidate-set}

All automated lens searching methods ultimately rely on visual
inspection to confirm the quality of potential lens candidates. The
neural networks provide a score representing a probability that a source
is a strong lens. The output of a probability score by neural networks, 
if provably consistent and robust, is of some value in an astronomy context 
as it allows a more fine-grained allocation of follow-up resources than 
the course-grained and highly stochastic "yes-no-maybe" grades produced
by human inspectors.

How to use this information to choose sources to
examine is up to the user. Our methodology involved examining the size
of candidate sets that satisfied various score criteria, and visually
inspecting several of these of a manageable size (i.e.~a few thousand).
Beyond this size, the law of diminishing returns makes visual inspection
less efficient as candidate sizes increase exponentially and the quality
of candidates decreases. Without a reference sample of lenses, it is
difficult to know what the optimal threshold for a candidate set is in
terms of the trade-off in purity and completeness.

We train networks with two different training sets (simulated
non-lenses, and real galaxies as non-lenses). We do this because real
galaxies that do not match the parameters of the simulated ETGs have a
high potential to confuse the network trained only on simulations; this
follows, as the network will have never seen anything resembling (say) a
spiral galaxy and thus its response to that morphology is undefined.
Beyond this intuition, the contribution of the two neural network scores
is a free parameter without real constraints. Inspection of candidate
sets of a similar size from each network suggests comparable purity.
To assist future searches, examining a much larger set of candidates, 
perhaps by citizen scientists, could assist in constraining the optimal settings.

In grading candidates, we discover many sources that could possibly
be a lens, where flux from a potential lensed source could be discerned
above the noise, and in a plausible configuration. However, these
sources are neither bright enough nor distinct enough to the human eye
to grade higher. These candidates, although not false positives in the
usual sense, may not be of a quality that warrants the expensive
spectroscopic follow-up required to do subsequent science. A future
training set could include simulated lenses with low signal-to-noise as
negative examples, to heighten the chance of activating on only the
strongest and most interesting discoverable lenses.

We reject many candidates offered by the lens-finder with high scores
due to insufficiently strong lensing features. However, we examine the
candidates as RGB images; the neural networks operate directly on the
calibrated FITS images and so are not as limited in dynamic range as the
human eye. We cannot be certain that the CNN is seeing something that
strongly indicates lensing that we cannot. This also suggests that
improvements in the tools used by the human vetters, such as a range of
contrast settings and single-band imaging or lens-subtracted images, may
improve the grading process.

\subsection{Comparison to other DES strong lens
searches}\label{comparison-to-other-des-strong-lens-searches}

Diehl et al \citeyearpar{DiehlBrightArcsSurvey2017} conducted a search
of the Dark Energy Survey science verification (SV) and Year 1 (Y1) observations and
identified 374 candidate strong lens systems of which the authors
designate 47 of high quality. 
The candidates were selected using several techniques including colour-based searches
(``Blue Near Anything'') and searches of a known catalogue of massive
Early-Type Galaxies. Assembling this candidate set required visual
inspection of approximately 400,000 cutout images. 
\citet[2018 in prep]{nord_observation_2016} 
searched DES 
SV and Y1 data for group and cluster-scale strong lenses, inspecting 250 square degrees of 
the SV and over 7000 catalogued clusters, identifying 53 lens candidates in 
the former and 46 in the latter, of which 21 were confirmed spectroscopically.
While the comparison
is complicated by the fact that our networks were trained specifically
for lenses at high redshift, we were able to obtain high completeness
after visual inspection of only \textasciitilde{}17,000 candidate images
(and only slightly less complete at 7,301). This suggests that our
neural network-based algorithm is considerably more efficient (in 
terms of human inspection time if not in terms of GPU resources).
This is consistent with the intuition that the morphological information learned
by the CNNs (but absent in the colour-based search methods) contains
information of high value in identifying strong lenses.

\subsection{Future work}\label{future-work}

The method detailed in this work is readily applicable to lens searches
at other redshifts and in other surveys. Improvements for future
searches will include expanding the variety of galaxies represented in
training sets, realistic variations in seeing in simulations, and
simulating lenses using models fit to real potential lens galaxies. The
number and architecture of the neural networks trained are still free
parameters. As more lenses are discovered in the survey these parameters
may be more easily constrained.

In this paper we have estimated completeness against lenses a human 
expert can confirm through visual inspection. Understanding the
detectability criterion better may enable the development of 
improved inspection tools or mechanisms such as displaying
lens-subtracted images. If the human thresholds are understood 
better training sets, that exclude real strong lenses that fall below this
threshold, will produce more useful candidate sets. Future work
will use simulations to better constrain the lensing parameters
that best facilitate human certainty. 

Realising the scientific potential of this catalogue will require
confirmation of the lenses, and the measurement of lens and source
redshifts. Higher-resolution imaging could also confirm lenses. With
improved seeing at or below \(.6\arcsec\), a robust measurement of the
Einstein radius would be possible, sufficient for mean total density
profile slope measurement using the method employed by Sonnenfeld
\citeyearpar{SonnenfeldSL2SGalaxyscaleLens2013} and others.

\section{Conclusion}\label{sec:conclusions}

Here we present a catalogue of 84 new high-quality strong lens
candidates from the Dark Energy Survey Year 3 coadd imaging. For our
target population of lenses at redshift \(> 0.8\)
in DES coadd images, we estimate this sample to include the majority
of those detectable in this imaging,
pending follow up spectroscopy to
confirm our candidates. If confirmation is forthcoming, this will
increase the sample of strong lenses at these distances by a factor of
3-5. To achieve this across the 5000 square degrees of the DES footprint
required only four to five hours of candidate inspection time by lens
experts.

In recent years, convolutional neural networks have proven a promising
technique in lens-finding and other astronomical classification
applications. Some tens of new candidate strong lenses have been
identified using deep learning already. With thousands or tens of
thousands waiting to be discovered in upcoming surveys, further
development of this method remains a promising area of research.

Here we apply convolutional neural networks to a search targeting lenses
at redshifts \(> 0.8\). The search is
motivated by the small sample of lenses known at these distances
(\(< 10\)) and the strong potential for a confirmed sample to impact our
understanding of the formation histories of elliptical galaxies at early
times, in particular by helping to constrain the evolution of the total
density slope with redshift. At the targeted redshift range, the lenses
have a particular morphology, where the central deflector is very faint 
in \(g\) band, which may be learned by the ANNs during training and 
reduce the number of false positives.

This method, and the pipeline developed in this work, can be readily
adapted to other surveys. Adjusting simulations to match the filters,
seeing and resolution of the target survey is likely necessary to
achieve good results. Future work will focus on increasing the purity of
samples further by discarding a greater proportion of false positive or
sub-optimal candidates. Our simulations can be improved, with more
realistic variations in colour and morphology (e.g.~groups,
mergers, or spiral galaxies) possible. 
The simulated seeing values were drawn from DES Y1 Science Verification values, and
was not matched to the DES Y3 coadd tiles used to construct the simulations. This
should be eliminated as a possible source of error.
Trained networks could also be
improved with online learning using the information gained by inspecting
candidates; this would complement e.g.~citizen science initiatives, with
human volunteers helping networks re-train by learning from false
positives labelled by expert inspection.

\emph{Acknowledgements}: 
This paper has gone through internal review by the DES collaboration. 

This research was supported by the Australian
Research Council Centre of Excellence for All Sky Astrophysics in 3
Dimensions (ASTRO 3D), through project number CE170100013.

TEC is supported by a Dennis Sciama Fellowship from the University of Portsmouth.

Funding for the DES Projects has been provided by the U.S. Department of Energy, the U.S. National Science Foundation, the Ministry of Science and Education of Spain, 
the Science and Technology Facilities Council of the United Kingdom, the Higher Education Funding Council for England, the National Center for Supercomputing 
Applications at the University of Illinois at Urbana-Champaign, the Kavli Institute of Cosmological Physics at the University of Chicago, 
the Center for Cosmology and Astro-Particle Physics at the Ohio State University,
the Mitchell Institute for Fundamental Physics and Astronomy at Texas A\&M University, Financiadora de Estudos e Projetos, 
Funda{\c c}{\~a}o Carlos Chagas Filho de Amparo à Pesquisa do Estado do Rio de Janeiro, Conselho Nacional de Desenvolvimento Cient{\'i}fico e Tecnol{\'o}gico and 
the Minist{\'e}rio da Ci{\^e}ncia, Tecnologia e Inova{\c c}{\~a}o, the Deutsche Forschungsgemeinschaft and the Collaborating Institutions in the Dark Energy Survey. 

The Collaborating Institutions are Argonne National Laboratory, the University of California at Santa Cruz, the University of Cambridge, Centro de Investigaciones Energ{\'e}ticas, 
Medioambientales y Tecnol{\'o}gicas-Madrid, the University of Chicago, University College London, the DES-Brazil Consortium, the University of Edinburgh, 
the Eidgen{\"o}ssische Technische Hochschule (ETH) Z{\"u}rich, 
Fermi National Accelerator Laboratory, the University of Illinois at Urbana-Champaign, the Institut de Ciències de l'Espai (IEEC/CSIC), 
the Institut de F{\'i}sica d'Altes Energies, Lawrence Berkeley National Laboratory, the Ludwig-Maximilians Universit{\"a}t M{\"u}nchen and the associated Excellence Cluster Universe, 
the University of Michigan, the National Optical Astronomy Observatory, the University of Nottingham, The Ohio State University, the University of Pennsylvania, the University of Portsmouth, 
SLAC National Accelerator Laboratory, Stanford University, the University of Sussex, Texas A\&M University, and the OzDES Membership Consortium.

Based in part on observations at Cerro Tololo Inter-American Observatory, National Optical Astronomy Observatory, which is operated by the Association of 
Universities for Research in Astronomy (AURA) under a cooperative agreement with the National Science Foundation.

The DES data management system is supported by the National Science Foundation under Grant Numbers AST-1138766 and AST-1536171.
The DES participants from Spanish institutions are partially supported by MINECO under grants AYA2015-71825, ESP2015-66861, FPA2015-68048, SEV-2016-0588, SEV-2016-0597, and MDM-2015-0509, 
some of which include ERDF funds from the European Union. IFAE is partially funded by the CERCA program of the Generalitat de Catalunya.
Research leading to these results has received funding from the European Research
Council under the European Union's Seventh Framework Program (FP7/2007-2013) including ERC grant agreements 240672, 291329, and 306478.
We  acknowledge support from the Australian Research Council Centre of Excellence for All-sky Astrophysics (CAASTRO), through project number CE110001020, and the Brazilian Instituto Nacional de Ci\^encia
e Tecnologia (INCT) e-Universe (CNPq grant 465376/2014-2).

This manuscript has been authored by Fermi Research Alliance, LLC under Contract No. DE-AC02-07CH11359 with the U.S. Department of Energy, Office of Science, Office of High Energy Physics. The United States Government retains and the publisher, by accepting the article for publication, acknowledges that the United States Government retains a non-exclusive, paid-up, irrevocable, world-wide license to publish or reproduce the published form of this manuscript, or allow others to do so, for United States Government purposes.
\hypertarget{tbl:new_candidates}{}
\onecolumn
\begin{center}
\begin{longtable}{@{}lllllll@{}}
\caption{\label{tbl:new_candidates}New candidates from visual inspection
of the neural network-selected sources, sorted by grade. }\\
\hline
Candidate & object id & RA & dec & grade & imag & \(z_{phot}\) \\
\hline
\endhead
DESJ0003-3348 & 139823797 & 0.8183 & -33.8012 & 3.00 & 19.77 & 0.56
\(\pm\) 0.31 \\
DESJ0347-2454 & 378100572 & 56.9356 & -24.9087 & 3.00 & 19.77 & 0.51
\(\pm\) 0.30 \\
DESJ0203-2338 & 67920213 & 30.7667 & -23.6340 & 3.00 & 19.15 & 0.58
\(\pm\) 0.31 \\
DESJ2216-4419 & 76102671 & 334.1592 & -44.3222 & 3.00 & 19.05 & 0.53
\(\pm\) 0.30 \\
DESJ2014-5757 & 166130477 & 303.5808 & -57.9504 & 2.67 & 20.64 & 0.76
\(\pm\) 0.34 \\
DESJ0143-0850 & 266637953 & 25.8622 & -8.8392 & 2.67 & 20.48 & 0.58
\(\pm\) 0.31 \\
DESJ0142-1831 & 266036534 & 25.7203 & -18.5211 & 2.67 & 19.62 & 0.57
\(\pm\) 0.31 \\
DESJ0124-1443 & 223066247 & 21.2211 & -14.7174 & 2.67 & 18.88 & 0.44
\(\pm\) 0.36 \\
DESJ0543-3752 & 443873820 & 85.7586 & -37.8770 & 2.67 & 20.06 & 0.54
\(\pm\) 0.30 \\
DESJ0415-4143 & 402556256 & 63.9363 & -41.7295 & 2.33 & 18.92 & 0.75
\(\pm\) 0.34 \\
DESJ0101-4917 & 290048397 & 15.4918 & -49.2939 & 2.33 & 20.31 & 0.68
\(\pm\) 0.33 \\
DESJ0357-5810 & 482065451 & 59.4035 & -58.1815 & 2.33 & 18.69 & 0.53
\(\pm\) 0.30 \\
DESJ0354-1609 & 386476783 & 58.5761 & -16.1645 & 2.33 & 19.34 & 0.53
\(\pm\) 0.30 \\
DESJ0212-0852 & 90442652 & 33.1051 & -8.8697 & 2.33 & 20.23 & 0.59
\(\pm\) 0.31 \\
DESJ0038-2936 & 157799078 & 9.6926 & -29.6019 & 2.33 & 21.13 & 0.71
\(\pm\) 0.34 \\
DESJ0058-5201 & 283879328 & 14.6447 & -52.0332 & 2.33 & 19.67 & 0.59
\(\pm\) 0.31 \\
DESJ0120-1820 & 354176405 & 20.1074 & -18.3338 & 2.33 & 20.77 & 0.71
\(\pm\) 0.34 \\
DESJ0305-1636 & 337847674 & 46.3197 & -16.6037 & 2.00 & 19.30 & 0.51
\(\pm\) 0.30 \\
DESJ2125-6504 & 191159999 & 321.3001 & -65.0741 & 2.00 & 19.88 & 0.73
\(\pm\) 0.34 \\
DESJ0024-3400 & 204184446 & 6.2373 & -34.0148 & 2.00 & 20.01 & 0.58
\(\pm\) 0.31 \\
DESJ0422-2132 & 496451011 & 65.5759 & -21.5461 & 2.00 & 20.23 & 0.54
\(\pm\) 0.30 \\
DESJ0327-3246 & 361760653 & 51.7973 & -32.7762 & 2.00 & 19.55 & 0.52
\(\pm\) 0.30 \\
DESJ0413-2344 & 400295190 & 63.4213 & -23.7395 & 2.00 & 20.17 & 0.58
\(\pm\) 0.31 \\
DESJ0412-1954 & 401080425 & 63.1615 & -19.9023 & 2.00 & 19.07 & 0.57
\(\pm\) 0.31 \\
DESJ0418-1817 & 405038616 & 64.6387 & -18.2982 & 2.00 & 22.13 & 0.86
\(\pm\) 0.36 \\
DESJ0508-2144 & 413900270 & 77.2053 & -21.7419 & 2.00 & 19.58 & 0.65
\(\pm\) 0.32 \\
DESJ2300-4454 & 106547800 & 345.0133 & -44.9065 & 1.67 & 20.01 & 0.51
\(\pm\) 0.34 \\
DESJ0546-2000 & 445925268 & 86.5211 & -20.0071 & 1.67 & 19.18 & 0.53
\(\pm\) 0.30 \\
DESJ2218-4504 & 75469120 & 334.7402 & -45.0738 & 1.67 & 19.83 & 0.55
\(\pm\) 0.30 \\
DESJ0109-0455 & 295037190 & 17.2945 & -4.9195 & 1.67 & 20.33 & 0.68
\(\pm\) 0.33 \\
DESJ0125-3645 & 266734513 & 21.2646 & -36.7664 & 1.67 & 19.99 & 0.59
\(\pm\) 0.31 \\
DESJ0332-1325 & 365125003 & 53.0106 & -13.4195 & 1.67 & 21.07 & 0.80
\(\pm\) 0.48 \\
DESJ0213-2413 & 90786519 & 33.2886 & -24.2292 & 1.67 & 21.35 & 0.65
\(\pm\) 0.32 \\
DESJ0557-2913 & 450317573 & 89.3717 & -29.2196 & 1.67 & 22.46 & 0.49
\(\pm\) 0.46 \\
DESJ0301-4426 & 337812631 & 45.4638 & -44.4405 & 1.67 & 21.29 & 0.75
\(\pm\) 0.38 \\
DESJ0313-3610 & 382872932 & 48.4060 & -36.1777 & 1.33 & 20.86 & 0.66
\(\pm\) 0.38 \\
DESJ0030-0426 & 207264051 & 7.6017 & -4.4478 & 1.33 & 20.25 & 0.63
\(\pm\) 0.32 \\
DESJ0530-6109 & 437004264 & 82.5136 & -61.1618 & 1.33 & 19.84 & 0.47
\(\pm\) 0.29 \\
DESJ0148-2251 & 254368847 & 27.1313 & -22.8577 & 1.33 & 19.64 & 0.58
\(\pm\) 0.31 \\
DESJ0110-2652 & 300525303 & 17.5715 & -26.8684 & 1.33 & 18.84 & 0.49
\(\pm\) 0.29 \\
DESJ0543-5825 & 446307824 & 85.7667 & -58.4196 & 1.33 & 21.47 & 0.66
\(\pm\) 0.40 \\
DESJ0533-2536 & 436520077 & 83.4555 & -25.6151 & 1.33 & 20.73 & 0.67
\(\pm\) 0.33 \\
DESJ0141-1303 & 264803099 & 25.2541 & -13.0509 & 1.33 & 20.74 & 0.63
\(\pm\) 0.32 \\
DESJ0521-5252 & 425857481 & 80.2896 & -52.8744 & 1.33 & 19.37 & 0.66
\(\pm\) 0.33 \\
DESJ0308-2106 & 343364859 & 47.2000 & -21.1039 & 1.33 & 22.28 & 0.77
\(\pm\) 0.36 \\
DESJ0332-5136 & 367575834 & 53.0118 & -51.6127 & 1.33 & 19.65 & 0.49
\(\pm\) 0.34 \\
DESJ0307-3444 & 342189632 & 46.8973 & -34.7414 & 1.33 & 20.28 & 0.55
\(\pm\) 0.30 \\
DESJ0202-2445 & 69413913 & 30.5277 & -24.7511 & 1.33 & 19.88 & 0.60
\(\pm\) 0.31 \\
DESJ0450-5044 & 483404421 & 72.5878 & -50.7436 & 1.33 & 20.80 & 0.55
\(\pm\) 0.30 \\
DESJ0440-5841 & 500132356 & 70.2452 & -58.6915 & 1.33 & 20.25 & 0.58
\(\pm\) 0.31 \\
DESJ0617-4033 & 464681328 & 94.3907 & -40.5590 & 1.00 & 20.71 & 0.49
\(\pm\) 0.29 \\
DESJ0535-5320 & 441369380 & 83.7508 & -53.3384 & 1.00 & 18.84 & 0.66
\(\pm\) 0.33 \\
DESJ0250-4104 & 324571256 & 42.6208 & -41.0717 & 1.00 & 19.37 & 0.55
\(\pm\) 0.30 \\
DESJ0226-3231 & 118076009 & 36.5643 & -32.5263 & 1.00 & 20.09 & 0.54
\(\pm\) 0.30 \\
DESJ2338-5101 & 138566300 & 354.5403 & -51.0208 & 1.00 & 20.42 & 0.58
\(\pm\) 0.31 \\
DESJ2245-4042 & 99179537 & 341.3828 & -40.7098 & 1.00 & 20.49 & 0.55
\(\pm\) 0.30 \\
DESJ0544-2509 & 443586921 & 86.0440 & -25.1584 & 1.00 & 19.89 & 0.46
\(\pm\) 0.29 \\
DESJ0400-2226 & 507569548 & 60.1166 & -22.4452 & 1.00 & 18.79 & 0.43
\(\pm\) 0.28 \\
DESJ0202-4105 & 68398953 & 30.6211 & -41.0887 & 1.00 & 19.67 & 0.66
\(\pm\) 0.33 \\
DESJ0624-4709 & 467288040 & 96.0659 & -47.1617 & 1.00 & 20.49 & 0.77
\(\pm\) 0.35 \\
DESJ0432-6002 & 470184935 & 68.2249 & -60.0451 & 1.00 & 20.62 & 0.71
\(\pm\) 0.34 \\
DESJ2222-5611 & 81574849 & 335.6330 & -56.1856 & 1.00 & 19.30 & 0.49
\(\pm\) 0.29 \\
DESJ0203-6322 & 66052645 & 30.8980 & -63.3693 & 1.00 & 19.42 & 0.53
\(\pm\) 0.30 \\
DESJ0603-4054 & 459178468 & 90.9654 & -40.9125 & 1.00 & 20.72 & 0.58
\(\pm\) 0.31 \\
DESJ0613-4509 & 464432181 & 93.3574 & -45.1528 & 1.00 & 20.30 & 0.61
\(\pm\) 0.32 \\
DESJ0546-4739 & 449145933 & 86.6012 & -47.6626 & 1.00 & 20.37 & 0.64
\(\pm\) 0.38 \\
DESJ0305-1024 & 341195944 & 46.2731 & -10.4032 & 1.00 & 21.10 & 0.63
\(\pm\) 0.32 \\
DESJ0150-0304 & 253888373 & 27.5379 & -3.0773 & 1.00 & 21.65 & 0.65
\(\pm\) 0.32 \\
DESJ0339-3914 & 373803496 & 54.8580 & -39.2375 & 1.00 & 20.11 & 0.53
\(\pm\) 0.30 \\
DESJ0038-2550 & 155609778 & 9.5932 & -25.8422 & 1.00 & 19.82 & 0.58
\(\pm\) 0.31 \\
DESJ2248-4955 & 101317774 & 342.2277 & -49.9234 & 1.00 & 19.35 & 0.49
\(\pm\) 0.29 \\
DESJ0315-2644 & 346529251 & 48.9752 & -26.7443 & 1.00 & 19.92 & 0.50
\(\pm\) 0.36 \\
DESJ2023-6457 & 163065099 & 305.8781 & -64.9653 & 1.00 & 19.98 & 0.51
\(\pm\) 0.30 \\
DESJ2337+0040 & 136806695 & 354.4976 & 0.6778 & 1.00 & 19.84 & 0.43
\(\pm\) 0.28 \\
DESJ0010-4315 & 182452355 & 2.6268 & -43.2541 & 1.00 & 20.65 & 0.79
\(\pm\) 0.35 \\
DESJ0408-2056 & 391106806 & 62.1010 & -20.9368 & 1.00 & 21.11 & 0.69
\(\pm\) 0.33 \\
DESJ0408-3956 & 390200758 & 62.1022 & -39.9407 & 1.00 & 20.20 & 0.54
\(\pm\) 0.30 \\
DESJ2352+0006 & 161118112 & 358.0487 & 0.1040 & 1.00 & 20.48 & 0.48
\(\pm\) 0.29 \\
DESJ0328-4714 & 364286007 & 52.1101 & -47.2339 & 1.00 & 22.57 & 0.71
\(\pm\) 0.38 \\
DESJ0155-1040 & 260575550 & 28.9336 & -10.6677 & 1.00 & 20.48 & 0.49
\(\pm\) 0.29 \\
DESJ2244-5903 & 97171633 & 341.0313 & -59.0510 & 1.00 & 18.82 & 0.50
\(\pm\) 0.30 \\
DESJ2319-5644 & 126893048 & 349.9322 & -56.7405 & 1.00 & 19.82 & 0.58
\(\pm\) 0.31 \\
DESJ0327-3312 & 364890268 & 51.9400 & -33.2036 & 1.00 & 21.19 & 0.72
\(\pm\) 0.42 \\
DESJ0314-2523 & 346534444 & 48.6681 & -25.3870 & 1.00 & 20.85 & 0.72
\(\pm\) 0.42 \\
\end{longtable}
\end{center}

\twocolumn

\bibliographystyle{mnras}
\renewcommand\refname{References}
\bibliography{./jacobs_des_highz_lenses.bib}

\begin{thebibliography}{}
\makeatletter
\relax
\def\mn@urlcharsother{\let\do\@makeother \do\$\do\&\do\#\do\^\do\_\do\%\do\~}
\def\mn@doi{\begingroup\mn@urlcharsother \@ifnextchar [ {\mn@doi@}
  {\mn@doi@[]}}
\def\mn@doi@[#1]#2{\def\@tempa{#1}\ifx\@tempa\@empty \href
  {http://dx.doi.org/#2} {doi:#2}\else \href {http://dx.doi.org/#2} {#1}\fi
  \endgroup}
\def\mn@eprint#1#2{\mn@eprint@#1:#2::\@nil}
\def\mn@eprint@arXiv#1{\href {http://arxiv.org/abs/#1} {{\tt arXiv:#1}}}
\def\mn@eprint@dblp#1{\href {http://dblp.uni-trier.de/rec/bibtex/#1.xml}
  {dblp:#1}}
\def\mn@eprint@#1:#2:#3:#4\@nil{\def\@tempa {#1}\def\@tempb {#2}\def\@tempc
  {#3}\ifx \@tempc \@empty \let \@tempc \@tempb \let \@tempb \@tempa \fi \ifx
  \@tempb \@empty \def\@tempb {arXiv}\fi \@ifundefined
  {mn@eprint@\@tempb}{\@tempb:\@tempc}{\expandafter \expandafter \csname
  mn@eprint@\@tempb\endcsname \expandafter{\@tempc}}}

\bibitem[\protect\citeauthoryear{Abbott et~al.,}{Abbott
  et~al.}{2018}]{AbbottDarkEnergySurvey2018}
Abbott T. M.~C.,  et~al., 2018, arXiv:1801.03181 [astro-ph]

\bibitem[\protect\citeauthoryear{Agnello, Kelly, Treu  \& Marshall}{Agnello
  et~al.}{2015}]{AgnelloDatamininggravitationally2015}
Agnello A.,  Kelly B.~C.,  Treu T.,   Marshall P.~J.,  2015, \mn@doi [MNRAS]
  {10.1093/mnras/stv037}, 448, 1446

\bibitem[\protect\citeauthoryear{Alard}{Alard}{2006}]{AlardAutomateddetectiongravitational2006}
Alard C.,  2006, arXiv:astro-ph/0606757

\bibitem[\protect\citeauthoryear{Amiaux et~al.,}{Amiaux
  et~al.}{2012}]{AmiauxEuclidMissionbuilding2012}
Amiaux J.,  et~al., 2012, \mn@doi [arXiv:1209.2228 [astro-ph]]
  {10.1117/12.926513}, p. 84420Z

\bibitem[\protect\citeauthoryear{Avestruz, Li, Lightman, Collett  \&
  Luo}{Avestruz et~al.}{2017}]{AvestruzAutomatedLensingLearner2017}
Avestruz C.,  Li N.,  Lightman M.,  Collett T.~E.,   Luo W.,  2017, preprint,
  1704, arXiv:1704.02322

\bibitem[\protect\citeauthoryear{Barnab{\`e}, Czoske, Koopmans, Treu  \&
  Bolton}{Barnab{\`e} et~al.}{2011}]{BarnabeTwodimensionalkinematicsSLACS2011}
Barnab{\`e} M.,  Czoske O.,  Koopmans L. V.~E.,  Treu T.,   Bolton A.~S.,
  2011, \mn@doi [MNRAS] {10.1111/j.1365-2966.2011.18842.x}, 415

\bibitem[\protect\citeauthoryear{Bellstedt et~al.,}{Bellstedt
  et~al.}{2018}]{BellstedtSLUGGSSurveycomparison2018a}
Bellstedt S.,  et~al., 2018, \mn@doi [MNRAS] {10.1093/mnras/sty456}

\bibitem[\protect\citeauthoryear{Bolton, Burles, Koopmans, Treu  \&
  Moustakas}{Bolton et~al.}{2006}]{BoltonSloanLensACS2006}
Bolton A.~S.,  Burles S.,  Koopmans L. V.~E.,  Treu T.,   Moustakas L.~A.,
  2006, \mn@doi [ApJ] {10.1086/498884}, 638, 703

\bibitem[\protect\citeauthoryear{Bonvin et~al.,}{Bonvin
  et~al.}{2016}]{BonvinH0LiCOWNewCOSMOGRAIL2016}
Bonvin V.,  et~al., 2016, \mn@doi [MNRAS] {10.1093/mnras/stw3006}, p. stw3006

\bibitem[\protect\citeauthoryear{Cao, Qin, Liu, Tsai  \& Li}{Cao
  et~al.}{2007}]{CaoLearningRankPairwise2007}
Cao Z.,  Qin T.,  Liu T.-Y.,  Tsai M.-F.,   Li H.,  2007, in Proceedings of the
  24th {{International Conference}} on {{Machine Learning}}. ICML '07.
{ACM}, New York, NY, USA, pp 129--136, \mn@doi{10.1145/1273496.1273513}

\bibitem[\protect\citeauthoryear{Cappellari et~al.,}{Cappellari
  et~al.}{2011}]{CappellariATLAS3Dprojectvolumelimited2011}
Cappellari M.,  et~al., 2011, \mn@doi [MNRAS]
  {10.1111/j.1365-2966.2010.18174.x}, 413, 813

\bibitem[\protect\citeauthoryear{Chan, Suyu, Chiueh, More, Marshall, Coupon,
  Oguri  \& Price}{Chan
  et~al.}{2015}]{ChanChitahStronggravitationallensHunter2015}
Chan J. H.~H.,  Suyu S.~H.,  Chiueh T.,  More A.,  Marshall P.~J.,  Coupon J.,
  Oguri M.,   Price P.,  2015, \mn@doi [ApJ] {10.1088/0004-637X/807/2/138}, 807

\bibitem[\protect\citeauthoryear{Choi, Park  \& Vogeley}{Choi
  et~al.}{2007}]{choiInternalCollectiveProperties2007a}
Choi Y.-Y.,  Park C.,   Vogeley M.~S.,  2007, \mn@doi [ApJ] {10.1086/511060},
  658, 884

\bibitem[\protect\citeauthoryear{{Chollet}}{{Chollet}}{2015}]{CholletKeras2015}
{Chollet} 2015, Keras

\bibitem[\protect\citeauthoryear{Collett}{Collett}{2015}]{collett_population_2015}
Collett T.~E.,  2015, \mn@doi [ApJ] {10.1088/0004-637X/811/1/20}, 811, 20

\bibitem[\protect\citeauthoryear{Collier, Smith  \& Lucey}{Collier
  et~al.}{2018}]{CollierImprovedmassconstraints2018}
Collier W.~P.,  Smith R.~J.,   Lucey J.~R.,  2018, \mn@doi [MNRAS]
  {10.1093/mnras/stx2297}, 473, 1103

\bibitem[\protect\citeauthoryear{Despali, Vegetti, White, Giocoli, Bosch  \&
  C}{Despali et~al.}{2018}]{DespaliModellinglineofsightcontribution2018}
Despali G.,  Vegetti S.,  White S. D.~M.,  Giocoli C.,  Bosch V.~D.,   C F.,
  2018, \mn@doi [Mon Not R Astron Soc] {10.1093/mnras/sty159}, 475, 5424

\bibitem[\protect\citeauthoryear{Diehl et~al.,}{Diehl
  et~al.}{2014}]{diehl_dark_2014}
Diehl H.~T.,  et~al., 2014. p. 91490V, \mn@doi{10.1117/12.2056982}, \url
  {http://adsabs.harvard.edu/abs/2014SPIE.9149E..0VD}

\bibitem[\protect\citeauthoryear{Diehl et~al.,}{Diehl
  et~al.}{2016}]{diehl_dark_2016}
Diehl H.~T.,  et~al., 2016, in Observatory {Operations}: {Strategies},
  {Processes}, and {Systems} {VI}. International Society for Optics and
  Photonics, p. 99101D, \mn@doi{10.1117/12.2233157}, \url
  {https://www.spiedigitallibrary.org/conference-proceedings-of-spie/9910/99101D/The-dark-energy-survey-and-operations--years-1-to/10.1117/12.2233157.short}

\bibitem[\protect\citeauthoryear{Diehl et~al.,}{Diehl
  et~al.}{2017}]{DiehlBrightArcsSurvey2017}
Diehl H.~T.,  et~al., 2017, \mn@doi [ApJS] {10.3847/1538-4365/aa8667}, 232, 15

\bibitem[\protect\citeauthoryear{Diehl et~al.}{Diehl
  et~al.}{2018}]{Diehl018jtu}
Diehl H.~T.,  et~al., 2018, \mn@doi [Proc. SPIE Int. Soc. Opt. Eng.]
  {10.1117/12.2312113}, 10704, 107040D

\bibitem[\protect\citeauthoryear{Ebeling, Stockmann, Richard, Zabl, Brammer,
  Toft  \& Man}{Ebeling
  et~al.}{2018}]{EbelingThirtyfoldExtremeGravitational2018}
Ebeling H.,  Stockmann M.,  Richard J.,  Zabl J.,  Brammer G.,  Toft S.,   Man
  A.,  2018, \mn@doi [ApJ Letters] {10.3847/2041-8213/aa9fee}, 852, L7

\bibitem[\protect\citeauthoryear{Einstein}{Einstein}{1936}]{EinsteinLenslikeactionstar1936}
Einstein A.,  1936, Science, 84, 506

\bibitem[\protect\citeauthoryear{Estrada et~al.,}{Estrada
  et~al.}{2007}]{EstradaSystematicSearchHigh2007}
Estrada J.,  et~al., 2007, \mn@doi [ApJ] {10.1086/512599}, 660, 1176

\bibitem[\protect\citeauthoryear{Fioc \& {Rocca-Volmerange}}{Fioc \&
  {Rocca-Volmerange}}{1999}]{FiocPEGASEmetallicityconsistentspectral1999a}
Fioc M.,  {Rocca-Volmerange} B.,  1999, arXiv:astro-ph/9912179

\bibitem[\protect\citeauthoryear{Flaugher et~al.,}{Flaugher
  et~al.}{2015}]{flaugher_dark_2015}
Flaugher B.,  et~al., 2015, \mn@doi [The Astronomical Journal]
  {10.1088/0004-6256/150/5/150}, 150, 150

\bibitem[\protect\citeauthoryear{Fukushima}{Fukushima}{1980}]{FukushimaNeocognitronselforganizingneural1980}
Fukushima K.,  1980, \mn@doi [Biol. Cybernetics] {10.1007/BF00344251}, 36, 193

\bibitem[\protect\citeauthoryear{Gavazzi, Marshall, Treu  \&
  Sonnenfeld}{Gavazzi et~al.}{2014}]{gavazzi_ringfinder:_2014}
Gavazzi R.,  Marshall P.~J.,  Treu T.,   Sonnenfeld A.,  2014, \mn@doi [ApJ]
  {10.1088/0004-637X/785/2/144}, 785, 144

\bibitem[\protect\citeauthoryear{Giacinto \& Roli}{Giacinto \&
  Roli}{2001}]{GiacintoDesigneffectiveneural2001}
Giacinto G.,  Roli F.,  2001, \mn@doi [Image and Vision Computing]
  {10.1016/S0262-8856(01)00045-2}, 19, 699

\bibitem[\protect\citeauthoryear{Guo, Liu, Oerlemans, Lao, Wu  \& Lew}{Guo
  et~al.}{2016}]{GuoDeeplearningvisual2016}
Guo Y.,  Liu Y.,  Oerlemans A.,  Lao S.,  Wu S.,   Lew M.~S.,  2016, \mn@doi
  [Neurocomputing] {10.1016/j.neucom.2015.09.116}, 187, 27

\bibitem[\protect\citeauthoryear{Hansen \& Salamon}{Hansen \&
  Salamon}{1990}]{HansenNeuralnetworkensembles1990}
Hansen L.~K.,  Salamon P.,  1990, \mn@doi [IEEE Transactions on Pattern
  Analysis and Machine Intelligence] {10.1109/34.58871}, 12, 993

\bibitem[\protect\citeauthoryear{He, Zhang, Ren  \& Sun}{He
  et~al.}{2016}]{HeDeepResidualLearning2016}
He K.,  Zhang X.,  Ren S.,   Sun J.,  2016. {Institute of Electrical and
  Electronics Engineers ( IEEE )}, Las Vegas, NV, pp 770--778

\bibitem[\protect\citeauthoryear{Hezaveh, Levasseur  \& Marshall}{Hezaveh
  et~al.}{2017}]{HezavehFastAutomatedAnalysis2017}
Hezaveh Y.~D.,  Levasseur L.~P.,   Marshall P.~J.,  2017

\bibitem[\protect\citeauthoryear{Hinton, Srivastava, Krizhevsky, Sutskever  \&
  Salakhutdinov}{Hinton et~al.}{2012}]{HintonImprovingneuralnetworks2012}
Hinton G.~E.,  Srivastava N.,  Krizhevsky A.,  Sutskever I.,   Salakhutdinov
  R.~R.,  2012, arXiv:1207.0580 [cs]

\bibitem[\protect\citeauthoryear{Hyde \& Bernardi}{Hyde \&
  Bernardi}{2009}]{Hydeluminositystellarmass2009}
Hyde J.~B.,  Bernardi M.,  2009, \mn@doi [MNRAS]
  {10.1111/j.1365-2966.2009.14783.x}, 396, 1171

\bibitem[\protect\citeauthoryear{Ilbert et~al.,}{Ilbert
  et~al.}{2009}]{ilbertCosmosPhotometricRedshifts2009}
Ilbert O.,  et~al., 2009, \mn@doi [ApJ] {10.1088/0004-637X/690/2/1236}, 690,
  1236

\bibitem[\protect\citeauthoryear{Ivezic et~al.,}{Ivezic
  et~al.}{2008}]{IvezicLSSTScienceDrivers2008}
Ivezic Z.,  et~al., 2008, arXiv:0805.2366 [astro-ph]

\bibitem[\protect\citeauthoryear{Jacobs, Glazebrook, Collett, More  \&
  McCarthy}{Jacobs et~al.}{2017}]{JacobsFindingstronglenses2017}
Jacobs C.,  Glazebrook K.,  Collett T.,  More A.,   McCarthy C.,  2017, \mn@doi
  [Mon Not R Astron Soc] {10.1093/mnras/stx1492}, 471, 167

\bibitem[\protect\citeauthoryear{Jordan \& Mitchell}{Jordan \&
  Mitchell}{2015}]{JordanMachinelearningtrends2015}
Jordan M.,  Mitchell T.,  2015, \mn@doi [Science] {10.1126/science.aaa8415},
  p.~255

\bibitem[\protect\citeauthoryear{Ju, Bibaut  \& {van der Laan}}{Ju
  et~al.}{2017}]{JuRelativePerformanceEnsemble2017}
Ju C.,  Bibaut A.,   {van der Laan} M.~J.,  2017, arXiv:1704.01664 [cs, stat]

\bibitem[\protect\citeauthoryear{Keeton}{Keeton}{2001}]{KeetonComputationalMethodsGravitational2001}
Keeton C.~R.,  2001, arXiv:astro-ph/0102340

\bibitem[\protect\citeauthoryear{Kelly et~al.,}{Kelly
  et~al.}{2017}]{Kellyindividualstarredshift2017}
Kelly P.~L.,  et~al., 2017, arXiv:1706.10279 [astro-ph]

\bibitem[\protect\citeauthoryear{Krizhevsky, Sutskever  \& Hinton}{Krizhevsky
  et~al.}{2012}]{krizhevsky_imagenet_2012}
Krizhevsky A.,  Sutskever I.,   Hinton G.~E.,  2012, in Pereira F.,  Burges C.
  J.~C.,  Bottou L.,   Weinberger K.~Q.,  eds, , Advances in {{Neural
  Information Processing Systems}} 25.
{Curran Associates, Inc.}, pp 1097--1105

\bibitem[\protect\citeauthoryear{Krogh \& Vedelsby}{Krogh \&
  Vedelsby}{1995}]{KroghNeuralnetworkensembles1995}
Krogh A.,  Vedelsby J.,  1995, in Advances in Neural Information Processing
  Systems. pp 231--238

\bibitem[\protect\citeauthoryear{Lanusse, Ma, Li, Collett, Li, Ravanbakhsh,
  Mandelbaum  \& Poczos}{Lanusse et~al.}{2017}]{LanusseCMUDeepLensDeep2017}
Lanusse F.,  Ma Q.,  Li N.,  Collett T.~E.,  Li C.-L.,  Ravanbakhsh S.,
  Mandelbaum R.,   Poczos B.,  2017, preprint, 1703, arXiv:1703.02642

\bibitem[\protect\citeauthoryear{LeCun, Boser, Denker, Henderson, Howard,
  Hubbard  \& Jackel}{LeCun
  et~al.}{1989}]{LeCunBackpropagationAppliedHandwritten1989}
LeCun Y.,  Boser B.,  Denker J.~S.,  Henderson D.,  Howard R.~E.,  Hubbard W.,
   Jackel L.~D.,  1989, \mn@doi [Neural Comput.] {10.1162/neco.1989.1.4.541},
  1, 541

\bibitem[\protect\citeauthoryear{LeCun, Bottou, Orr  \& M{\"u}ller}{LeCun
  et~al.}{1998}]{LeCunEfficientBackProp1998}
LeCun Y.,  Bottou L.,  Orr G.~B.,   M{\"u}ller K.-R.,  1998, in Orr G.~B.,
  M{\"u}ller K.-R.,  eds, Lecture Notes in Computer ScienceNo.~1524, Neural
  {{Networks}}: {{Tricks}} of the {{Trade}}.
{Springer Berlin Heidelberg}, pp 9--50, \mn@doi{10.1007/3-540-49430-8_2}

\bibitem[\protect\citeauthoryear{Lenzen, Schindler  \& Scherzer}{Lenzen
  et~al.}{2004}]{LenzenAutomaticdetectionarcs2004}
Lenzen F.,  Schindler S.,   Scherzer O.,  2004, \mn@doi [A&A]
  {10.1051/0004-6361:20034619}, 416, 11

\bibitem[\protect\citeauthoryear{Li, Frenk, Cole, Gao, Bose  \& Hellwing}{Li
  et~al.}{2016}]{LiConstraintsidentitydark2016}
Li R.,  Frenk C.~S.,  Cole S.,  Gao L.,  Bose S.,   Hellwing W.~A.,  2016,
  \mn@doi [MNRAS] {10.1093/mnras/stw939}, 460

\bibitem[\protect\citeauthoryear{Marshall, Hogg, Moustakas, Fassnacht, Brada{\v
  c}, {Tim Schrabback}  \& Blandford}{Marshall
  et~al.}{2009}]{marshall_automated_2009}
Marshall P.~J.,  Hogg D.~W.,  Moustakas L.~A.,  Fassnacht C.~D.,  Brada{\v c}
  M.,  {Tim Schrabback}  Blandford R.~D.,  2009, \mn@doi [ApJ]
  {10.1088/0004-637X/694/2/924}, 694, 924

\bibitem[\protect\citeauthoryear{Marshall et~al.,}{Marshall
  et~al.}{2016}]{MarshallSPACEWARPSCrowdsourcing2016}
Marshall P.~J.,  et~al., 2016, \mn@doi [MNRAS] {10.1093/mnras/stv2009}, 455

\bibitem[\protect\citeauthoryear{Metcalf et~al.,}{Metcalf
  et~al.}{2018}]{MetcalfStrongGravitationalLens2018}
Metcalf R.~B.,  et~al., 2018, arXiv:1802.03609 [astro-ph]

\bibitem[\protect\citeauthoryear{More, Cabanac, More, Alard, Limousin, Kneib,
  Gavazzi  \& Motta}{More et~al.}{2012}]{MoreCFHTLSStrongLensing2012}
More A.,  Cabanac R.,  More S.,  Alard C.,  Limousin M.,  Kneib J.-P.,  Gavazzi
  R.,   Motta V.,  2012, \mn@doi [ApJ] {10.1088/0004-637X/749/1/38}, 749, 38

\bibitem[\protect\citeauthoryear{More et~al.,}{More
  et~al.}{2016}]{MoreSPACEWARPSII2016}
More A.,  et~al., 2016, \mn@doi [MNRAS] {10.1093/mnras/stv1965}, 455

\bibitem[\protect\citeauthoryear{Morganson et~al.,}{Morganson
  et~al.}{2018}]{MorgansonDarkEnergySurvey2018}
Morganson E.,  et~al., 2018, \mn@doi [Publications of the Astronomical Society
  of the Pacific] {10.1088/1538-3873/aab4ef}, 130, 074501

\bibitem[\protect\citeauthoryear{Newton, Marshall, Treu, Auger, Gavazzi,
  Bolton, Koopmans  \& Moustakas}{Newton et~al.}{2011}]{NewtonSloanLensACS2011}
Newton E.~R.,  Marshall P.~J.,  Treu T.,  Auger M.~W.,  Gavazzi R.,  Bolton
  A.~S.,  Koopmans L. V.~E.,   Moustakas L.~A.,  2011, \mn@doi [ApJ]
  {10.1088/0004-637X/734/2/104}, 734

\bibitem[\protect\citeauthoryear{Nguyen, Yosinski  \& Clune}{Nguyen
  et~al.}{2015}]{NguyenDeepneuralnetworks2015}
Nguyen A.,  Yosinski J.,   Clune J.,  2015, in Proceedings of the {{IEEE
  Conference}} on {{Computer Vision}} and {{Pattern Recognition}}. pp 427--436

\bibitem[\protect\citeauthoryear{Nord et~al.,}{Nord
  et~al.}{2016}]{nord_observation_2016}
Nord B.,  et~al., 2016, \mn@doi [ApJ] {10.3847/0004-637X/827/1/51}, 827, 51

\bibitem[\protect\citeauthoryear{Oldham \& Auger}{Oldham \&
  Auger}{2018}]{OldhamGalaxystructuremultiple2018}
Oldham L.,  Auger M.,  2018, \mn@doi [MNRAS] {10.1093/mnras/stx2969}, 474, 4169

\bibitem[\protect\citeauthoryear{Petrillo et~al.,}{Petrillo
  et~al.}{2017}]{PetrilloFindingStrongGravitational2017}
Petrillo C.~E.,  et~al., 2017, arXiv:1702.07675 [astro-ph]

\bibitem[\protect\citeauthoryear{Quider, Pettini, Shapley  \& Steidel}{Quider
  et~al.}{2009}]{Quiderultravioletspectrumgravitationally2009}
Quider A.~M.,  Pettini M.,  Shapley A.~E.,   Steidel C.~C.,  2009, \mn@doi
  [MNRAS] {10.1111/j.1365-2966.2009.15234.x}, 398

\bibitem[\protect\citeauthoryear{Refaeilzadeh, Tang  \& Liu}{Refaeilzadeh
  et~al.}{2009}]{RefaeilzadehCrossValidation2009}
Refaeilzadeh P.,  Tang L.,   Liu H.,  2009, in , Encyclopedia of {{Database
  Systems}}.
{Springer, Boston, MA}, pp 532--538, \mn@doi{10.1007/978-0-387-39940-9_565}

\bibitem[\protect\citeauthoryear{Remus, Dolag, Naab, Burkert, Hirschmann,
  Hoffmann  \& Johansson}{Remus
  et~al.}{2017}]{RemusCoEvolutionTotalDensity2017}
Remus R.-S.,  Dolag K.,  Naab T.,  Burkert A.,  Hirschmann M.,  Hoffmann T.~L.,
    Johansson P.~H.,  2017, \mn@doi [MNRAS] {10.1093/mnras/stw2594}, 464, 3742

\bibitem[\protect\citeauthoryear{Rosenblatt}{Rosenblatt}{1957}]{RosenblattPerceptronaperceivingrecognizing1957}
Rosenblatt F.,  1957, Cornell Aeronautical Lab

\bibitem[\protect\citeauthoryear{Ruff, Gavazzi, Marshall, Treu, Auger  \&
  Brault}{Ruff et~al.}{2011}]{RuffSL2SGalaxyscaleLens2011}
Ruff A.~J.,  Gavazzi R.,  Marshall P.~J.,  Treu T.,  Auger M.~W.,   Brault F.,
  2011, \mn@doi [ApJ] {10.1088/0004-637X/727/2/96}, 727, 96

\bibitem[\protect\citeauthoryear{Schmidhuber}{Schmidhuber}{2015}]{SchmidhuberDeeplearningneural2015}
Schmidhuber J.,  2015, \mn@doi [Neural Networks]
  {10.1016/j.neunet.2014.09.003}, 61, 85

\bibitem[\protect\citeauthoryear{Seidel \& Bartelmann}{Seidel \&
  Bartelmann}{2007}]{seidel_arcfinder:_2007}
Seidel G.,  Bartelmann M.,  2007, \mn@doi [A&A] {10.1051/0004-6361:20066097},
  472, 12

\bibitem[\protect\citeauthoryear{Shankar et~al.,}{Shankar
  et~al.}{2018}]{ShankarRevisitingbulgehalo2018}
Shankar F.,  et~al., 2018, \mn@doi [Mon Not R Astron Soc]
  {10.1093/mnras/stx3086}, 475, 2878

\bibitem[\protect\citeauthoryear{Sonnenfeld, Treu, Gavazzi, Suyu, Marshall,
  Auger  \& Nipoti}{Sonnenfeld
  et~al.}{2013}]{SonnenfeldSL2SGalaxyscaleLens2013}
Sonnenfeld A.,  Treu T.,  Gavazzi R.,  Suyu S.~H.,  Marshall P.~J.,  Auger
  M.~W.,   Nipoti C.,  2013, \mn@doi [ApJ] {10.1088/0004-637X/777/2/98}, 777,
  98

\bibitem[\protect\citeauthoryear{Stark, Swinbank, Ellis, Dye, Smail  \&
  Richard}{Stark et~al.}{2008}]{Starkformationassemblytypical2008}
Stark D.~P.,  Swinbank A.~M.,  Ellis R.~S.,  Dye S.,  Smail I.~R.,   Richard
  J.,  2008, \mn@doi [Nature] {10.1038/nature07294}, 455, 775

\bibitem[\protect\citeauthoryear{Team et~al.,}{Team
  et~al.}{2016}]{TheTheanoDevelopmentTeamTheanoPythonframework2016}
Team T. T.~D.,  et~al., 2016, arXiv:1605.02688 [cs]

\bibitem[\protect\citeauthoryear{{The DES Collaboration}}{{The DES
  Collaboration}}{2005}]{TheDESCollaborationDarkEnergySurvey2005}
{The DES Collaboration} 2005, arXiv:astro-ph/0510346

\bibitem[\protect\citeauthoryear{{Tim de Zeeuw} et~al.,}{{Tim de Zeeuw}
  et~al.}{2002}]{TimdeZeeuwSAURONprojectII2002}
{Tim de Zeeuw} P.,  et~al., 2002, \mn@doi [Mon Not R Astron Soc]
  {10.1046/j.1365-8711.2002.05059.x}, 329, 513

\bibitem[\protect\citeauthoryear{Treu}{Treu}{2010}]{treu_strong_2010}
Treu T.,  2010, \mn@doi [ARA&A] {10.1146/annurev-astro-081309-130924}, 48, 87

\bibitem[\protect\citeauthoryear{Treu \& Koopmans}{Treu \&
  Koopmans}{2004}]{TreuMassiveDarkMatter2004}
Treu T.,  Koopmans L. V.~E.,  2004, \mn@doi [ApJ] {10.1086/422245}, 611, 739

\bibitem[\protect\citeauthoryear{Vegetti, Lagattuta, McKean, Auger, Fassnacht
  \& Koopmans}{Vegetti et~al.}{2012}]{VegettiGravitationaldetectionlowmass2012}
Vegetti S.,  Lagattuta D.~J.,  McKean J.~P.,  Auger M.~W.,  Fassnacht C.~D.,
  Koopmans L. V.~E.,  2012, \mn@doi [Nature] {10.1038/nature10669}, 481, 341

\bibitem[\protect\citeauthoryear{Walsh, Carswell  \& Weymann}{Walsh
  et~al.}{1979}]{Walsh0957561twin1979}
Walsh D.,  Carswell R.~F.,   Weymann R.~J.,  1979, \mn@doi [Nature]
  {10.1038/279381a0}, 279, 381

\bibitem[\protect\citeauthoryear{Weijmans, Krajnovi{\'c}, {van de Ven},
  Oosterloo, Morganti, Zeeuw  \& T}{Weijmans
  et~al.}{2008}]{Weijmansshapedarkmatter2008}
Weijmans A.-M.,  Krajnovi{\'c} D.,  {van de Ven} G.,  Oosterloo T.~A.,
  Morganti R.,  Zeeuw D.,   T P.,  2008, \mn@doi [MNRAS]
  {10.1111/j.1365-2966.2007.12680.x}, 383, 1343

\bibitem[\protect\citeauthoryear{Zheng et~al.,}{Zheng
  et~al.}{2012}]{Zhengmagnifiedyounggalaxy2012}
Zheng W.,  et~al., 2012, \mn@doi [Nature] {10.1038/nature11446}, 489, 406

\bibitem[\protect\citeauthoryear{Zwicky}{Zwicky}{1937}]{ZwickyNebulaeGravitationalLenses1937}
Zwicky F.,  1937, \mn@doi [Phys. Rev.] {10.1103/PhysRev.51.290}, 51, 290

\makeatother
\end{thebibliography}

\section*
{Affiliations}

$^{1}$ Centre for Astrophysics \& Supercomputing, Swinburne University of Technology, Victoria 3122, Australia\\
$^{2}$ ARC Centre of Excellence for All Sky Astrophysics in 3 Dimensions (ASTRO 3D)\\
$^{3}$ Institute of Cosmology \& Gravitation, University of Portsmouth, Portsmouth, PO1 3FX, UK\\
$^{4}$ School of Software and Electrical Engineering, Swinburne University of Technology, P.O. Box 218, Hawthorn, VIC 3122, Australia\\
$^{5}$ Cerro Tololo Inter-American Observatory, National Optical Astronomy Observatory, Casilla 603, La Serena, Chile\\
$^{6}$ Department of Physics \& Astronomy, University College London, Gower Street, London, WC1E 6BT, UK\\
$^{7}$ Department of Physics and Electronics, Rhodes University, PO Box 94, Grahamstown, 6140, South Africa\\
$^{8}$ Fermi National Accelerator Laboratory, P. O. Box 500, Batavia, IL 60510, USA\\
$^{9}$ LSST, 933 North Cherry Avenue, Tucson, AZ 85721, USA\\
$^{10}$ CNRS, UMR 7095, Institut d'Astrophysique de Paris, F-75014, Paris, France\\
$^{11}$ Sorbonne Universit\'es, UPMC Univ Paris 06, UMR 7095, Institut d'Astrophysique de Paris, F-75014, Paris, France\\
$^{12}$ Kavli Institute for Particle Astrophysics \& Cosmology, P. O. Box 2450, Stanford University, Stanford, CA 94305, USA\\
$^{13}$ SLAC National Accelerator Laboratory, Menlo Park, CA 94025, USA\\
$^{14}$ Laborat\'orio Interinstitucional de e-Astronomia - LIneA, Rua Gal. Jos\'e Cristino 77, Rio de Janeiro, RJ - 20921-400, Brazil\\
$^{15}$ Observat\'orio Nacional, Rua Gal. Jos\'e Cristino 77, Rio de Janeiro, RJ - 20921-400, Brazil\\
$^{16}$ Department of Astronomy, University of Illinois at Urbana-Champaign, 1002 W. Green Street, Urbana, IL 61801, USA\\
$^{17}$ National Center for Supercomputing Applications, 1205 West Clark St., Urbana, IL 61801, USA\\
$^{18}$ Institut de F\'{\i}sica d'Altes Energies (IFAE), The Barcelona Institute of Science and Technology, Campus UAB, 08193 Bellaterra (Barcelona) Spain\\
$^{19}$ Centro de Investigaciones Energ\'eticas, Medioambientales y Tecnol\'ogicas (CIEMAT), Madrid, Spain\\
$^{20}$ Department of Physics, IIT Hyderabad, Kandi, Telangana 502285, India\\
$^{21}$ Department of Astronomy/Steward Observatory, 933 North Cherry Avenue, Tucson, AZ 85721-0065, USA\\
$^{22}$ Jet Propulsion Laboratory, California Institute of Technology, 4800 Oak Grove Dr., Pasadena, CA 91109, USA\\
$^{23}$ Kavli Institute for Cosmological Physics, University of Chicago, Chicago, IL 60637, USA\\
$^{24}$ Instituto de Fisica Teorica UAM/CSIC, Universidad Autonoma de Madrid, 28049 Madrid, Spain\\
$^{25}$ Institut d'Estudis Espacials de Catalunya (IEEC), 08193 Barcelona, Spain\\
$^{26}$ Institute of Space Sciences (ICE, CSIC),  Campus UAB, Carrer de Can Magrans, s/n,  08193 Barcelona, Spain\\
$^{27}$ Department of Astronomy, University of Michigan, Ann Arbor, MI 48109, USA\\
$^{28}$ Department of Physics, University of Michigan, Ann Arbor, MI 48109, USA\\
$^{29}$ Department of Astronomy, University of California, Berkeley,  501 Campbell Hall, Berkeley, CA 94720, USA\\
$^{30}$ Lawrence Berkeley National Laboratory, 1 Cyclotron Road, Berkeley, CA 94720, USA\\
$^{31}$ Department of Physics, ETH Zurich, Wolfgang-Pauli-Strasse 16, CH-8093 Zurich, Switzerland\\
$^{32}$ Santa Cruz Institute for Particle Physics, Santa Cruz, CA 95064, USA\\
$^{33}$ Center for Cosmology and Astro-Particle Physics, The Ohio State University, Columbus, OH 43210, USA\\
$^{34}$ Department of Physics, The Ohio State University, Columbus, OH 43210, USA\\
$^{35}$ Max Planck Institute for Extraterrestrial Physics, Giessenbachstrasse, 85748 Garching, Germany\\
$^{36}$ Universit\"ats-Sternwarte, Fakult\"at f\"ur Physik, Ludwig-Maximilians Universit\"at M\"unchen, Scheinerstr. 1, 81679 M\"unchen, Germany\\
$^{37}$ Harvard-Smithsonian Center for Astrophysics, Cambridge, MA 02138, USA\\
$^{38}$ Australian Astronomical Observatory, North Ryde, NSW 2113, Australia\\
$^{39}$ Departamento de F\'isica Matem\'atica, Instituto de F\'isica, Universidade de S\~ao Paulo, CP 66318, S\~ao Paulo, SP, 05314-970, Brazil\\
$^{40}$ Department of Astronomy, The Ohio State University, Columbus, OH 43210, USA\\
$^{41}$ Instituci\'o Catalana de Recerca i Estudis Avan\c{c}ats, E-08010 Barcelona, Spain\\
$^{42}$ School of Physics and Astronomy, University of Southampton,  Southampton, SO17 1BJ, UK\\
$^{43}$ Brandeis University, Physics Department, 415 South Street, Waltham MA 02453\\
$^{44}$ Instituto de F\'isica Gleb Wataghin, Universidade Estadual de Campinas, 13083-859, Campinas, SP, Brazil\\
$^{45}$ Computer Science and Mathematics Division, Oak Ridge National Laboratory, Oak Ridge, TN 37831\\
$^{46}$ Argonne National Laboratory, 9700 South Cass Avenue, Lemont, IL 60439, USA\\
$^{47}$ Institute for Astronomy, University of Edinburgh, Edinburgh EH9 3HJ, UK\\

\appendix

\onecolumn
\section{Keras model summary}
\begin{longtable}{@{}lll@{}}
\caption{\label{tbl:keras_model}Output of the Keras model summary for the convolutional neural networks used in this lens search.}\\
\hline
Layer (type) & Output Shape & Param count \\
\hline
conv2d\_13 (Conv2D) & (None, 96, 50, 50) & 34944 \\
max\_pooling2d\_10 (MaxPooling & (None, 96, 24, 24) & 0 \\
conv2d\_14 (Conv2D) & (None, 128, 24, 24) & 307328 \\
activation\_19 (Activation) & (None, 128, 24, 24) & 0 \\
max\_pooling2d\_11 (MaxPooling & (None, 128, 11, 11) & 0 \\
conv2d\_15 (Conv2D) & (None, 256, 11, 11) & 295168 \\
activation\_20 (Activation) & (None, 256, 11, 11) & 0 \\
conv2d\_16 (Conv2D) & (None, 256, 11, 11) & 590080 \\
dropout\_13 (Dropout) & (None, 256, 11, 11) & 0 \\
activation\_21 (Activation) & (None, 256, 11, 11) & 0 \\
max\_pooling2d\_12 (MaxPooling & (None, 256, 5, 5) & 0 \\
dropout\_14 (Dropout) & (None, 256, 5, 5) & 0 \\
flatten\_4 (Flatten) & (None, 6400) & 0 \\
dense\_10 (Dense) & (None, 1024) & 6554624 \\
activation\_22 (Activation) & (None, 1024) & 0 \\
dropout\_15 (Dropout) & (None, 1024) & 0 \\
dense\_11 (Dense) & (None, 1024) & 1049600 \\
activation\_23 (Activation) & (None, 1024) & 0 \\
dropout\_16 (Dropout)       & (None, 1024) & 0 \\
dense\_12 (Dense)           & (None, 2) & 2050 \\
activation\_24 (Activation) & (None, 2) & 0 \\

\hline
Total params: 8,833,794 & & \\
Trainable params: 8,833,794 & & \\
Non-trainable params: 0 & & \\
\hline
\end{longtable}
\twocolumn


\label{lastpage}
\end{document}